\documentclass[12pt,preprint]{aastex}

\slugcomment{To appear in ApJ}

\shorttitle{NGC~1333 from {\it Spitzer}}
\shortauthors{Gutermuth et al.}

\begin{document}


\title{{\it Spitzer} Observations of NGC 1333: A Study of Structure and Evolution in a Nearby Embedded Cluster}


\author{R. A. Gutermuth\altaffilmark{1}, P. C. Myers\altaffilmark{1}, S. T. Megeath\altaffilmark{2}, L. E. Allen\altaffilmark{1}, J. L. Pipher\altaffilmark{3}, J. Muzerolle\altaffilmark{4}, A. Porras\altaffilmark{5}, E. Winston\altaffilmark{1}, G. Fazio\altaffilmark{1}}


\altaffiltext{1}{Smithsonian Astrophysical Observatory}
\altaffiltext{2}{University of Toledo}
\altaffiltext{3}{University of Rochester}
\altaffiltext{4}{Steward Observatory}
\altaffiltext{5}{INAOE Mexico}


\begin{abstract}
We present a comprehensive analysis of structure in the young, embedded cluster, NGC~1333 using members identified with {\it Spitzer} and 2MASS photometry based on their IR-excess emission.
In total, 137 members are identified in this way, composed of 39 protostars and 98 more evolved pre-main sequence stars with disks.
Of the latter class, four are transition/debris disk candidates.
The fraction of exposed pre-main sequence stars with disks is $83\% \pm 11\%$, showing that there is a measurable diskless pre-main sequence population.
The sources in each of the Class~I and Class~II evolutionary states are shown to have very different spatial distributions relative to the distribution of the dense gas in their natal cloud.
However, the distribution of nearest neighbor spacings among these two groups of sources are found to be quite similar, with a strong peak at spacings of 0.045~pc.
Radial and azimuthal density profiles and surface density maps computed from the identified YSOs show that NGC~1333 is elongated and not strongly centrally concentrated, confirming previous claims in the literature.
We interpret these new results as signs of a low velocity dispersion, extremely young cluster that is not in virial equilibrium. 


\end{abstract}





\section{Introduction}

Observations of embedded, star-forming clusters and groups show that the stellar distributions are often elongated, clumpy, or both \citep[cf.][]{ll03,gute05,teix06,alle07}, and the structure seems tied to the distribution of dense gas in the clusters' natal molecular clouds.
Most clouds have some non-spherical structure, yet current results suggest that the associated clusters have varying degrees of agreement with the cloud's structure depending on how deeply the members are embedded \citep{gute05,teix06}.
Given the high frequency of asymmetric structure in clouds, it seems reasonable to assume that the exposed, relatively structureless clusters we observe may have been asymmetric in a previous epoch.
With the ejection of the majority of their natal gas and adequate time to migrate from their birth sites, the imprint of the underlying cloud structure could be lost rather quickly in a recently exposed young cluster.
Thus the structure we measure in a distribution of young stellar objects (YSO) relative to the dense gas in the associated cloud may be a reasonable proxy for the current dynamical state of an embedded cluster.


The launch of {\it Spitzer} has provided a potent new facility for studies of the structure of young and embedded clusters.
While some members of these young clusters are diskless pre-main sequence stars (Class~III), the majority of the membership are sources with excess emission at mid-IR wavelengths \citep{hll01}, made up of a mixture of protostars (Class~I), still embedded and accreting from dense spherical envelopes, and the slightly more evolved pre-main sequence stars with circumstellar disks (Class~II).
The mid-IR spectral energy distributions (SED) of these objects are dominated by the emission from their dusty circumstellar material, making them easily distinguishable from pure photospheric sources such as unrelated field stars and indistinguishable diskless cluster members.
{\it Spitzer}'s sensitivity to mid-IR emission makes it the best tool currently available for identifying {\it and characterizing} YSOs with IR-excess, and that sensitivity is sufficient to detect these sources down to the Hydrogen--burning mass limit for regions within the nearest kiloparsec \citep{gute04}.
The complete census of YSOs with disks in a young cluster represents a high-confidence sample of bona fide cluster members, and for the youngest regions, such a sample includes a high fraction of the total number of members \citep{hll01}.
Furthermore, by separating the two canonical evolutionary classes of YSOs that {\it Spitzer} so effectively detects, we are able to probe both the recent overall star-forming activity in the region as traced by the stars with disks, and the immediate star formation as represented by the protostars, since this phase is expected to be short-lived relative to the former. 

Many young clusters are dominated by heavy and spatially variable extinction from dust within their natal molecular cloud environment.  
Another advantage {\it Spitzer} brings to studies of these regions is that the mid-IR wavelengths targeted by {\it Spitzer}'s imaging instruments (Infrared Array Camera, or IRAC, at 3.6-8.0~$\mu$m and Multiband Imaging Photometer for {\it Spitzer}, or MIPS, at 24-160~$\mu$m) are less affected by extinction from dust in comparison to near-IR or visible wavelengths \citep[e.g.][]{ccm89}.
Recent papers have done an excellent job of characterizing the reddening law in the IRAC bandpasses \citep{inde05,flah07,huar07} and have made a first attempt at doing the same for the 24~$\mu$m channel of MIPS \citep{flah07,huar07}.  
Given the wealth of information {\it Spitzer} can bring to the study of embedded clusters, we have surveyed over thirty clustered star-forming regions within the nearest kiloparsec through the Guaranteed Time Observations (GTO) program for the IRAC and MIPS instrument teams.
With these data, we can not only provide a full census of the YSOs with IR-excess in each region, but also perform a detailed examination and comparative analysis of structure in young, embedded clusters.  

NGC~1333 has been a popular target for observations of deeply embedded protostars via radio wavelengths due to the presence of an unprecendented number of molecular outflows \citep[e.g.][]{ks00} associated with several bright IRAS sources considered Class~0 protostars \citep{jenn87}, all in relatively close proximity to the Sun \citep[250~pc,][]{enoc06}.
Many of the outflows are traced by shock-induced emission, a clear sign that they are indeed affecting the local, quiescent cloud material \citep[e.g.][]{wala05}.
Because of this, some studies have claimed that the NGC~1333 dense molecular cloud core is in the process of being destroyed by influence from outflows \citep[e.g.][]{wari96,sk01,quil05}.
In addition to the numerous protostars, there is a cluster of pre-main sequence stars identified by number counts analysis in the near-IR \citep{svs76,asr94,lal96,wilk04}.
\citet{lal96} suggested that the distribution was well-described as a ``double cluster'', having two distinct surface density maxima.

Here we present a {\it Spitzer} IRAC and MIPS imaging and photometric analysis of the NGC~1333 young cluster, one of the most nearby large membership ($N>100$) clusters in the {\it Spitzer} Young Cluster Survey.
We achieve a near-complete census of the cluster membership that possesses circumstellar material, significantly surpassing the sensitivity of ground-based mid-IR surveys \citep[cf.][and references therein.]{rebu03}.  
In addition, we statistically infer the population of pre-main sequence stars that lack disks, enabling an estimate of the fraction of members with disks.
Finally, using the identified YSOs, we apply both established and recently developed methods for characterizing the structure of this cluster, in specific reference to previous claims of structure found in the literature.    



\section{Observations\label{obs}}


\subsection{Multi-band Point Source Photometry}

\subsubsection{IRAC Imaging}

Observations of the NGC~1333 region were obtained with the IRAC instrument \citep{fazi04} onboard {\it Spitzer} \citep{wern04} over two epochs: February~10,~2004 (PID~6) and September~8,~2004 (PID~178). 
The total area covered by all four IRAC bandpasses in the final mosaic is roughly $40' \times 30'$, centered on RA(J2000) = $03^h 29^m 00^s$, Dec.(J2000) = $31^d 20^m 00^s$.
The typical total exposure time per band is 41.6~seconds, comprised of four 10.4~second integrations.
High Dynamic Range (HDR) mode was used, so there are matching images at 0.4 second integration times to obtain photometry for bright sources.
Mosaics were built at the native instrument resolution of 1.2'' per pixel with the standard Basic Calibrated Data (BCD) products provided by the {\it Spitzer} Science Center from their standard data processing pipeline version S13.2.
Additional processing of bright source artifacts (``jailbar'', ``pulldown'', ``muxbleed'', and ``banding'') is performed on these data via customized versions of Interactive Data Language (IDL) scripts developed by the IRAC instrument team \citep{hora04,piph04}.
Mosaicking was performed using Gutermuth's WCSmosaic IDL package, which integrates the following features: a redundancy-based cosmic ray detection and rejection algorithm; frame-by-frame distortion correction, derotation, and subpixel offsetting in a single transformation (to minimize data smoothing); and on-the-fly background matching.
WCSmosaic is built with heavy dependence on the FITS and WCS access and manipulation procedures provided by the IDL Astronomy Users Library \citep{land93}.

Point source detection and aperture photometry is performed via Gutermuth's IDL photometry and visualization tool, PhotVis \citep{gute04}, version 1.10.  
Of significant note in this new version of PhotVis is the inclusion of a spatially variable source detection threshold made possible through the computation of a median absolute deviation uncertainty map from the Gaussian-convolved image computed as part of the DAOfind detection algorithm \citep{stet87}.  
With this modification, the improved algorithm is able to more clearly distinguish point sources from background variations, dramatically improving source detection robustness in regions of bright, structured nebulosity common in IRAC images of active star-forming regions.
The full-width at half-maximum (FWHM) of every detection is measured and all detections with FWHM~$>$~3.6'' are considered resolved and removed.
Some marginally resolved but highly structured shock emission sources were also visually identified and removed.  
Aperture photometry was performed on all sources with an aperture radius of 2.4'' and background flux estimation was done within concentric sky annuli of inner and outer radii of 2.4'' and 7.2'' respectively, following the method of previous work \citep{mege04,gute04}.
Photometric zero points were derived directly from the calibrations presented in \citet{reac05}, using standard aperture corrections for the radii adopted. 
See Table~\ref{photpar} for 90\% differential completeness limits and zero-point magnitude calibration values adopted for all bandpasses and all data sources. 
PhotVis is built with heavy dependence on the FITS access and manipulation procedures and DAOphot version 1 port to IDL provided by the IDL Astronomy Users Library \citep{land93}.

\subsubsection{MIPS Imaging}

MIPS \citep{riek04} imaging of NGC~1333 was obtained on February 3, 2004 (PID 58) at the medium scan rate.
Twelve scans of 1.75 degrees at 160'' offsets were needed to cover the full IRAC coverage.  
Image treatment and mosaicking was performed with the MIPS instrument team's Data Analysis Tool (DAT).
The DAT is used to treat bright source artifacts, persistence artifacts, and cosmic ray hits in each image.
It then mosaics the images together, removing spatial scale distortion before integrating each frame into the mosaic at the native MIPS resolution of 2.54'' per pixel.
Due to the low resolution and sensitivity of the 70~$\mu$m and 160~$\mu$m data obtained simultaneously, only the 24~$\mu$m data are presented and used in this paper.  
An analysis of these data for the entire Perseus molecular cloud is presented in \citet{rebu07} as part of the c2d Legacy Survey \citep{evan03}.

MIPS 24~$\mu$m point source detection and aperture photometry were also performed with PhotVis version 1.10.
Aperture and inner and outer sky annulus radii were chosen to be 7.6'', 7.6'', and 17.8'', respectively.
The photometric zero point of $[24]=14.6$ for a 1~DN/s flux measurement with the adopted aperture and sky annulus radii was chosen to make the mean $[5.8]-[24]$ colors of the nine photospheric sources detected with MIPS equal to zero (reduced from thirteen sources after applying iterative statistics techniques). 
RMS scatter of the internal measurement is 0.04~magnitude.

\subsubsection{2MASS $JHK_S$ Photometry}

All sources detected by {\it Spitzer} are matched to Two-Micron All Sky Survey (2MASS) Point Source Catalog (PSC) sources, extending the potential wavelength coverage for these objects to J band at 1.2~$\mu$m.
The NGC~1333 {\it Spitzer} survey region is also one of the locations covered by deeper 2MASS observations, and the merged data from these deeper observations were kindly provided by Roc Cutri ahead of public release.
The deeper data is reduced using the 2MASS final release data pipeline, which is documented in detail in \citet{skru06}.
These data are taken from observations of six times the integration time of the standard 2MASS observations as part of the 2MASS extended mission, yielding approximately a 1~magnitude improvement in the data completeness limit and reducing measurement uncertainty for sources within a magnitude of the detection limit of the regular 2MASS data \citep{chs01}.

\subsubsection{The Final Catalog} 

Source list matching for the final catalog is performed in stages.
First, the four IRAC band source lists are matched using Gutermuth's WCSphotmatch utility, with a maximum 1'' radial tolerance among matches. 
Second, the 2MASS catalog is integrated with the same software, using the mean positions of all entries in the merged IRAC catalog.
The RMS scatter of the radial residual offsets between each IRAC band and 2MASS ranges from 330 to 390 milliarcseconds.
Finally, the MIPS 24~$\mu$m catalog is integrated with WCSphotmatch using a wider 3'' maximum radial tolerance, now using the mean 2MASS and IRAC positions for all entries in the second stage catalog.
The RMS scatter of the radial residual offsets between the MIPS catalog and 2MASS is 750 milliarcseconds.

\subsection{Molecular Cloud Maps}

Tracers of the distribution of gas in a molecular cloud have varying levels of reliability, depending on environment, and that of NGC~1333 is especially challenging.   
Many recent papers have shown that CO and its isotopes are relatively poor column density tracers \citep[e.g.][]{kram99} outside of a rather narrow range of column densities (A. Goodman, priv. comm.), and the many active outflows in NGC~1333 add considerable confusion. 
The high galactic latitude of NGC~1333 has low field star surface density and a general lack of significant numbers of background giants compared to regions like the galactic bulge \citep[e.g. B68,][]{all01}, thus extinction maps derived from the reddened near-IR colors of background sources are generally low resolution.
In addition, the high overall extinction ($A_V > 10$) in NGC~1333 and the presence of a partially embedded population of pre-main sequence stars will bias the map toward lower column densities, particularly where cluster members are more numerous than field stars \citep{gute05}.
Both cold (e.g. in the cold dust filament where the Class~0 objects IRAS~4a, 4b, and 4c are forming) and hot (e.g. in the northern reflection nebulosity illuminated by late B stars) environments are found in NGC~1333, making conversion of submillimeter dust continuum emission to gas column density particularly uncertain \citep{schn05}.
Recent work with other dense gas tracers holds promise, such as the $N_2H^+$ map presented in \cite{wals07}, but this particular study (11'~$\times$~11') does not cover the large field of view of the {\it Spitzer} imaging data. 

Given the many pitfalls, we adopt two different cloud mapping methods here based only on their utility for two specific functions: tracing large scale, low-level extinction reliably, and tracing dense gas {\it morphology} reliably.  
For the former purpose, we create extinction maps using the $H-K_S$ colors of field stars from the combined 2MASS data.  
The method used is identical to that presented in Section~3.1 of \citet{gute05}, based on elements of both the NICE \citep{lada94} and NICER \citep{la01} algorithms.  
The resulting maps are sensitive to extinctions from 1~$A_V$ to 18~$A_V$ and have variable resolution based on local stellar surface densities, ranging from 70'' to 200''.
By characterizing low-level extinction well, we are able to customize our color- and magnitude--based cuts to mitigate extragalactic contamination to utilize the fact that extinction shifts the number density of contaminators per magnitude to dimmer magnitudes (see Section~\ref{iracclass} below).
For tracing the morphology of dense gas, we adopt the JCMT/SCUBA 850~$\mu$m map built from the combination of all such observations of this region found in the JCMT online archive \citep{kirk07}.
The bulk of these observations were originally published and analyzed in \citet{sk01} with expansion of the original field presented in \citet{hatc05}.
While derived column densities from this map are highly uncertain, 850~$\mu$m flux should trace the overall morphology of the densest portion of the cloud with reasonable accuracy to enable a comparison with the spatial distribution of cluster members (see Section~\ref{distro1} below).





\section{The {\it Spitzer} View of NGC~1333}

\begin{figure}
\epsscale{1}
\plotone{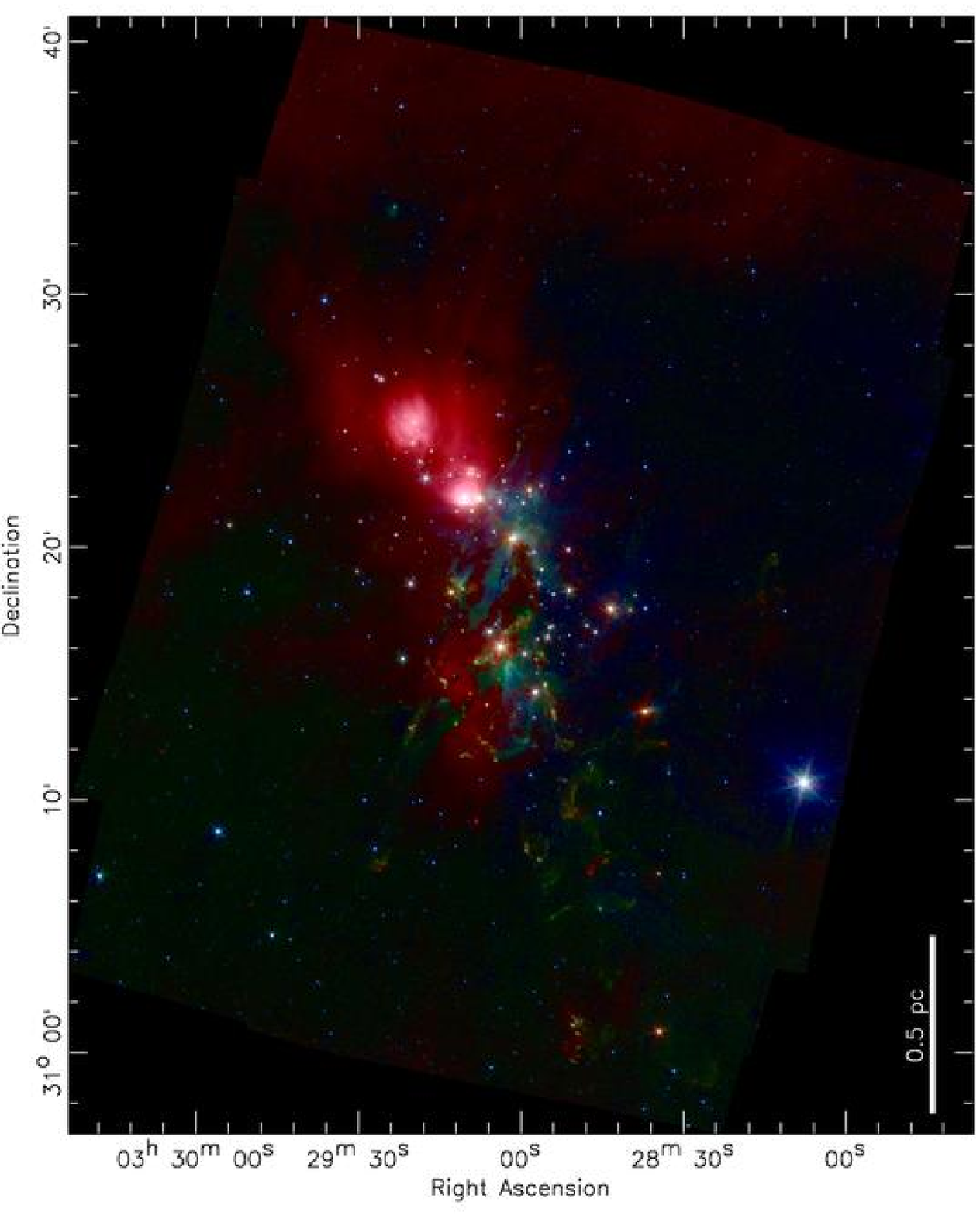}
\caption{Color-composite image of the IRAC mosaics of NGC~1333; 3.6, 4.5, and 8.0~$\mu$m images are mapped to blue, green, and red, respectively.\label{pretty1}}
\end{figure}

\begin{figure}
\epsscale{1}
\plotone{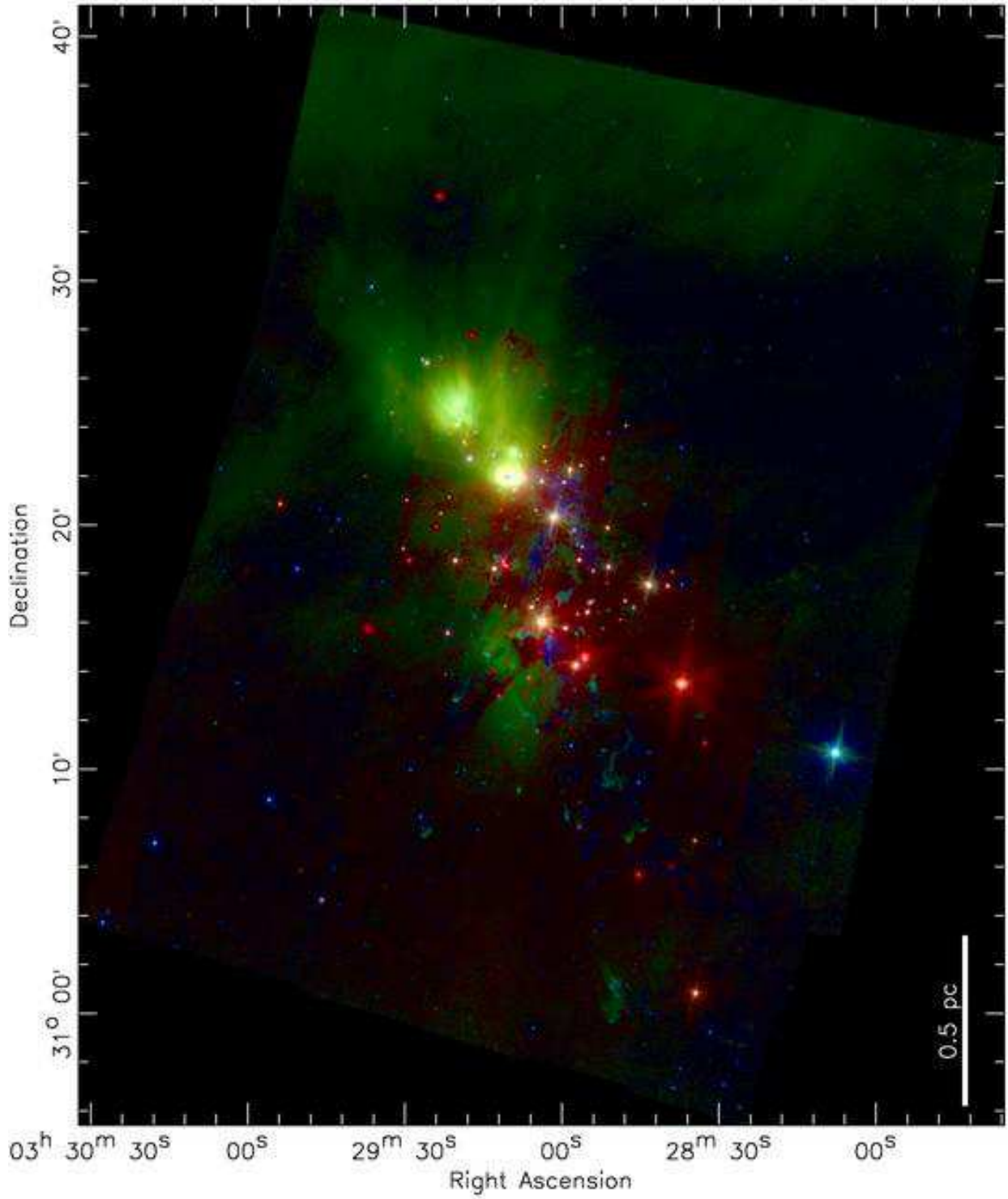}
\caption{Color-composite image of the IRAC and MIPS mosaics of NGC~1333; 4.5, 8.0, and 24~$\mu$m images are mapped to blue, green, and red, respectively.\label{pretty2}}
\end{figure}

We present color renditions of the {\it Spitzer} IRAC and MIPS mosaics in Figures~\ref{pretty1}~\&~\ref{pretty2}\footnote{For reference, the near-IR view presented in Fig.~16 of \citet{ll03} is the deepest near-IR image of the entire main cluster region currently available, and many of the region's well-studied objects are marked in Fig.~1 of \citet{wals07}.}.
In Figure~\ref{pretty1}, we have mapped the IRAC 3.6, 4.5, and 8.0~$\mu$m mosaics in blue, green, and red respectively.
NGC~1333 is unique in the Young Cluster Survey in that it is dominated by a complex network of 4.5~$\mu$m-dominated (dark green) shock emission from several different outflows colliding with ambient cloud material.
A more diffuse, blue-green color shows regions of scattered light from the surfaces of the molecular material.
Finally, the bright red region in the north coincides with the visibly exposed reflection nebulosity that is ``NGC~1333''.
The 8.0~$\mu$m emission in red here is commonly considered to be UV-excited polycyclic aromatic hydrocarbon (PAH) feature emission, where the UV source is usually one or more intermediate or high-mass stars.
There is also a less intense, but similarly sized region of PAH emission in the south end of the region, with the IRAS~4abc molecular cloud filament seen in absorption (RA = 03:29:07, Dec. = +31:13:30). 
This is a new feature revealed by the extinction-penetrating capabilities and mid-IR sensitivity of {\it Spitzer}, likely a bubble being evacuated by a previously unknown massive star.  
We have identified a candidate for the exciting star, ASR~54, located near the projected center of the newly revealed PAH nebulosity structure and apparently behind the dense, dusty molecular cloud filament.
While not detected in the 2MASS dataset presented here, ASR~54 was detected at $K = 15.74$ by \citet{asr94}.   
Given the extinction law of \citet{flah07}, the IRAC photometry of ASR~54 is consistent with a heavily extinguished photosphere ($A_K\sim10$).  
As such, for an assumed distance of 250~pc, we estimate that the absolute $K$-band magnitude of ASR~54 is $K = -1.25$, consistent with a late B-type star. 
The 24~$\mu$m photometry of ASR~54 also suggests there is some excess emission in this bandpass, resulting in this source's inclusion as a transition/debris disk candidate (see below).

In Figure~\ref{pretty2}, we have mapped the IRAC 4.5, 8.0, and MIPS 24~$\mu$m mosaics in blue, green, and red respectively.
In these colors, the many protostars in this region are made visually identifiable by the bright red halo around each source.  
The pre-main sequence stars with disks and diskless stars are noticeably bluer in this view.
The bright red source on the east side of the cluster in this image (RA = 03:29:37.1, Dec. = +31:15:46) is detected only at 24~$\mu$m.  
Given the elongated nature of the detection, no detections in any other IRAC or MIPS bands, and its position offset from the molecular cloud core, we are confident that this source is an asteroid. 

\section{The Census of YSOs in NGC~1333}


Here we describe the empirically-derived methods used to classify sources with infrared excess in NGC~1333.  
We have adopted a three phase approach to achieve maximum sensitivity to YSOs in all evolutionary states that exhibit IR-excess under a wide range of nebulosity environments from the combined multiband photometric dataset.
In each phase independently, we account for and attempt to mitigate contamination from extragalactic sources that masquerade as bona fide YSOs in the field.















\subsection{IRAC Four-Band Source Characterization\label{iracclass}}

As our primary dataset, we consider the set of all sources with detections ($\sigma < 0.2$~mag) in all four IRAC bandpasses, a total of 578 sources.
There are now several methods in the literature to identify and classify YSOs with such photometry \citep[cf.][]{alle04,mege04,lada06}.
Each of the cited schemes depends in large part on the 3.6~$\mu$m photometry to distinguish between YSOs of differing evolutionary class.
However, in deeply embedded clusters such as NGC~1333, YSO classification schemes that rely on 3.6~$\mu$m photometry will suffer to some degree because of the degeneracy caused by reddening from dust.
It has been shown that the extinction law flattens considerably through the 4.5-8.0~$\mu$m regime, only rising in the 3.6~$\mu$m bandpass \citep{lutz99}, and this has been confirmed in several {\it Spitzer}-based studies \citep{inde05,flah07,huar07}. 
To dramatically reduce the risk that differential extinction will confuse our classification of YSOs, we have adopted a new scheme that relies primarily on $[4.5]-[5.8]$ color to distinguish between Class~I and Class~II YSOs.  
This color is much less affected by dust extinction than any $[3.6]$-based color, yet it can still be employed to distinguish between Class~I and Class~II YSOs \citep[see Fig.~2 of ][]{hart05}.

\begin{figure}
\epsscale{.6}
\plotone{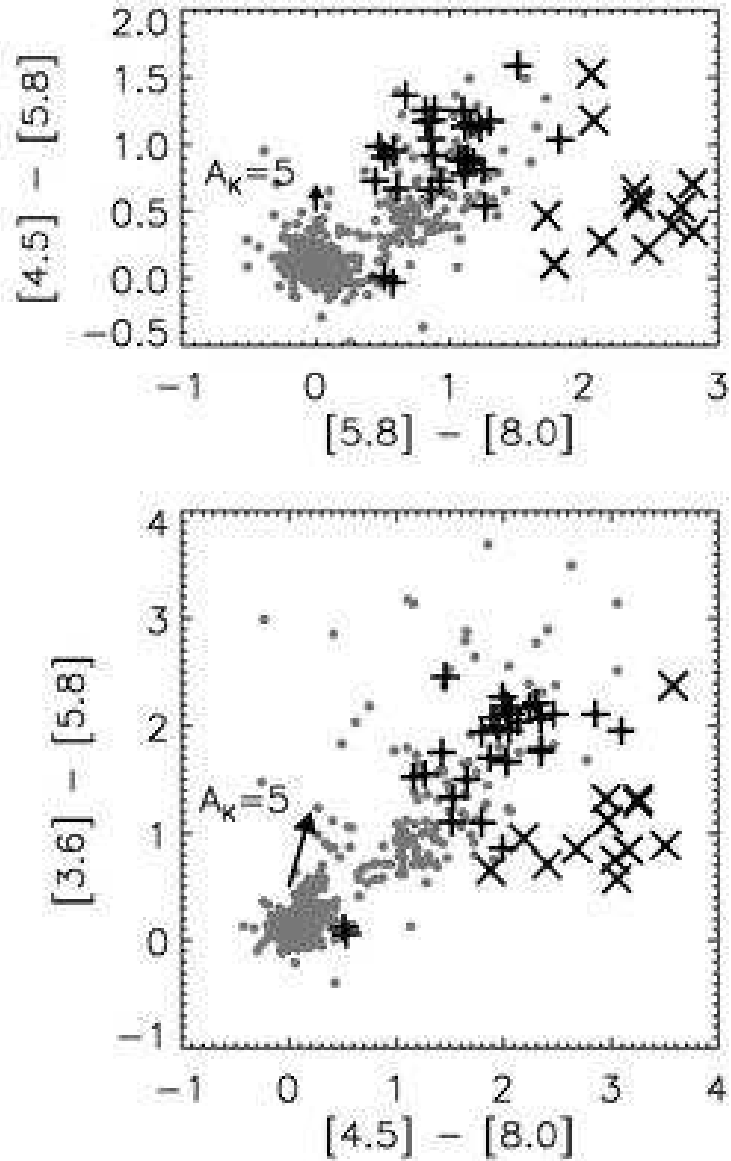}
\caption{Color-color diagrams used for identifying PAH emission sources.  The upper plot is the $[4.5]-[5.8]$ vs. $[5.8]-[8.0]$ color-color diagram, and the lower one is the $[4.5]-[8.0]$ vs. $[3.6]-[5.8]$ color-color diagram.  PAH-emission sources are marked with crosses, and broad-line AGN are marked with plusses.\label{pah1}}
\end{figure}

\begin{figure}
\epsscale{1}
\plotone{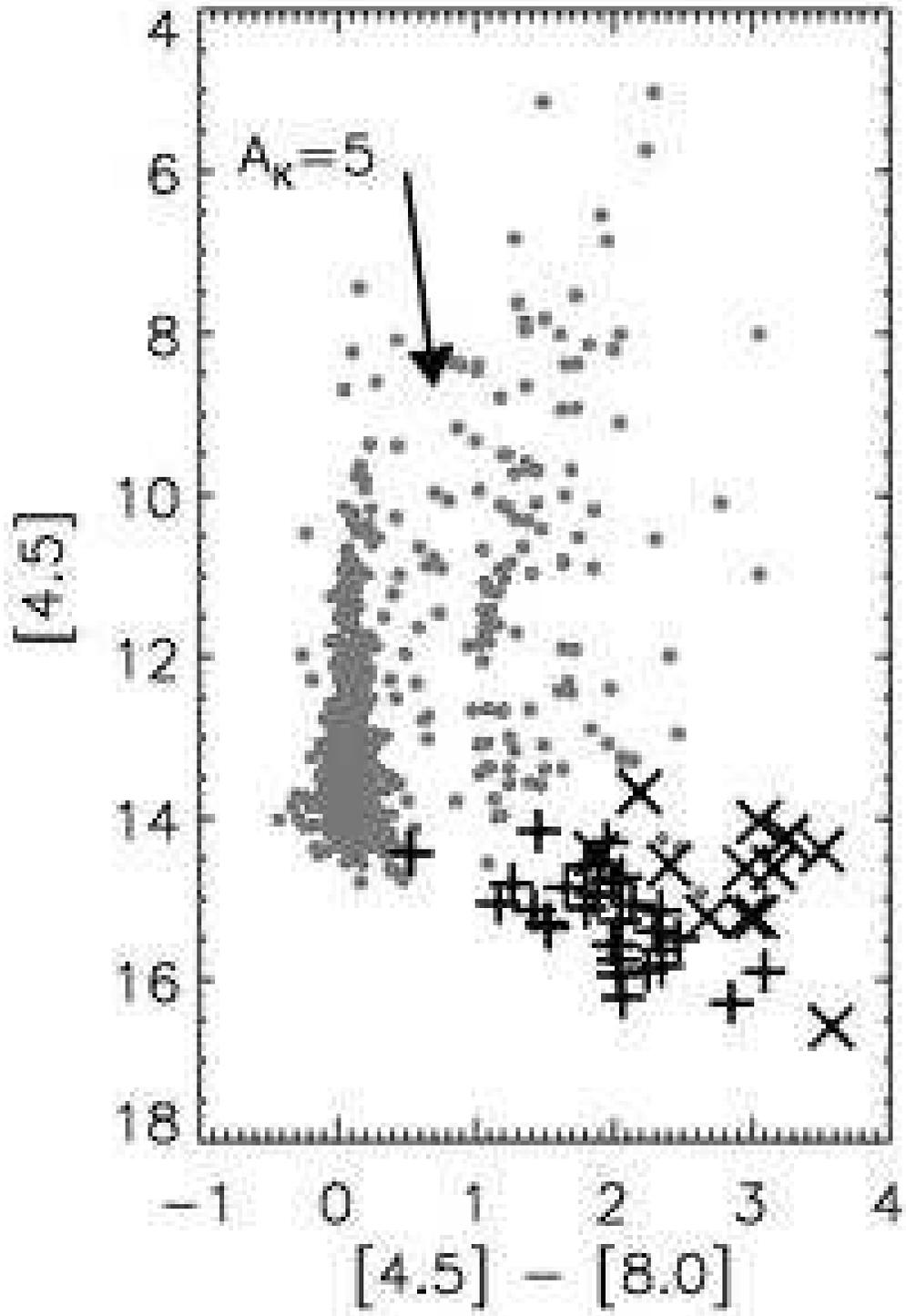}
\caption{$[4.5]$ vs. $[4.5]-[8.0]$ color-magnitude diagram, used for identifying likely broad-line AGN.   PAH-emission sources are marked with crosses, and broad-line AGN are marked with plusses.\label{agn1}}
\end{figure}

Before attempting to identify YSOs from these data, we must first remove various extragalactic contaminants that may be misidentified as YSOs.  
Initially, we identify and remove 13 sources dominated by PAH feature emission, likely star-forming galaxies and weak-line active galactic nuclei (AGN), following the method described in Appendix~\ref{bootes} (see the color-color diagrams in Figure~\ref{pah1}).
We then identify and remove 31 sources that fall within the region of color-magnitude space dominated by broad-line AGN (see Figure~\ref{agn1}).  
It is important to note here that the presence of a molecular cloud introduces a bias against a source of a given magnitude being an AGN, as extinction from dust in the cloud shifts the density of extragalactic sources per magnitude toward dimmer magnitudes.  
To account for this, we artificially assume that every source is located behind the cloud and deredden their photometry using the $A_K$ value from our extinction map at each source's position.  
Then we apply the AGN cut in $[4.5]$ vs. $[4.5]-[8.0]$ color-magnitude space described in Appendix~\ref{bootes} to the {\it dereddened} photometry.
Those sources flagged in this way have colors entirely consistent with low luminosity YSOs, thus we will reexamine these sources for signs of a protostellar signature using the MIPS 24~$\mu$m photometry in Section~\ref{mipsphase} below.
See the left plot of Figure~\ref{ysomap} for the spatial distribution of sources filtered from the NGC~1333 source list as likely contaminants.

To test the filtration process, we compare the resulting distribution of PAH emission and AGN candidate sources with models derived from our re-reduction of the {\it Spitzer} survey for galaxies in the Bootes field (see Appendix~\ref{bootes}).
These models are built from the $[5.8]$ magnitude distribution shown in Appendix~\ref{bootes}, scaled for the area of the NGC~1333 IRAC four-channel coverage and adjusted to account for the relative areas of differing extinction via the 2MASS-derived extinction map and the IRAC reddening law of \cite{flah07}.
See Figure~\ref{vermhist} for comparison histograms for the PAH and AGN sources.
Overall, the agreement between the filtered source distribution and the models is very good, suggesting that very few bona fide YSOs were mistaken as contaminants and discarded. 
However, there is likely residual contamination of $1 \pm 1$ sources after application of the contaminant filtration process.
According to the Bootes characterization, there is a higher probability that these are mistaken as Class~I rather than Class~II sources (see Appendix~\ref{bootes}).

We now attempt to filter one final contaminant from the source list before we proceed with identifying and classifying the YSO population: unresolved blobs of shocked emission from high-velocity outflows interacting with the cold molecular cloud.
These sources have a particularly large 4.5~$\mu$m excess due to shocked molecular hydrogen line emission in this band \citep{smit06}, and this fact is exploited to pull these sources out of the final list.
As was the case for the list of likely AGN, the shock emission source list can hold some bona fide protostars.
Hence we also will check this list for MIPS detections in Section~\ref{mipsphase}. 
NGC~1333 has the most structured shock emission of any cluster in the {\it Spitzer} Young Cluster Survey, and thus we have used it to empirically determine the region of IRAC color space that these sources occupy.
Based on our findings, all sources with photometry that obeys all of the following constraints are likely dominated by shock emission and thus are removed:

\begin{displaymath}
[3.6]-[4.5] > \frac{1.2}{0.55} \times (([4.5]-[5.8])-0.3) + 0.8
\end{displaymath}
\begin{displaymath}
[4.5]-[5.8] \le 0.85
\end{displaymath}
\begin{displaymath}
[3.6]-[4.5] > 1.05 
\end{displaymath}

Four sources are removed using these contraints.
The remaining sources are dominated by stellar sources, and infrared excess sources still present are true YSOs of both Class~II and Class~I evolutionary states.  
As mentioned above, we have adopted the $[4.5]-[5.8]$ color as the primary discriminant between protostars and stars with disks in this empirically determined classification scheme.
It has been shown in the low extinction environment of the nearby Taurus molecular cloud that YSOs can be successfully classified with this color \citep{hart05}, and the reddening effect of dust along the line of sight is much less severe for this color than for $[3.6]-[4.5]$ \citep[e.g.][]{flah07}.
Sources are likely protostars if they have extremely red discriminant color ($[4.5]-[5.8] > 1$).  
Additionally, any sources with moderately red discriminant color ($0.7 < [4.5]-[5.8] \le 1.0$) that also have $[3.6]-[4.5] > 0.7$ are likely protostars, though in rare cases, a highly reddened Class~II source could have these colors as well. 

Once the protostars are removed from consideration, we move on to identifying the Class~II sources from those that remain.  
All sources with the following color constraints are likely Class~II sources:

\begin{displaymath}
[4.5]-[8.0] > 0.5
\end{displaymath}
\begin{displaymath}
[3.6]-[5.8] > 0.35
\end{displaymath}
\begin{displaymath}
[3.6]-[5.8] \le \frac{0.14}{0.04} \times (([4.5]-[8.0])-0.5) + 0.5
\end{displaymath}

In total, we identify 29 sources with protostellar colors and 93 with colors consistent with a pre-main sequence star with a circumstellar disk using this method.
All other sources still remaining are consistent with photospheric colors.
Almost all of these objects are field stars unrelated to the cluster itself.  
Unfortunately, pre-main sequence stars that lack disks but are bona fide members of the cluster are still mixed in with this group and cannot be identified from the data presented here.
For this work, we simply account for these sources statistically, using star counting methods similar to those presented in \citet{gute05} (See Section~\ref{df} below).

This classification scheme has been verified using the $JHK_S$ and IRAC photometry for the nearby, low-extinction Taurus molecular cloud YSOs presented in \citet{hart05}.
Of the 82 Taurus sources presented in that work, 72 have full four band IRAC photometric coverage with $\sigma < 0.2$~mag.
Our method classifies as Class~III all 24 sources considered Class~III in that work.
Of the 41 classified as Class~II in \citet{hart05}, our method classifies 39 as Class~II and 2 as Class~III.
Finally, we classify as Class~I 4 of the 6 sources called Class~0, Class~I, or Class~I/II in \cite{hart05}; the other two we classify as Class~II. 

The discrepant sources in each evolutionary class have measurably weaker IR-excess than the non-discrepant ones, as measured by the $\alpha_{IRAC}$ classification described in \citet{lada06}.
In that work, the slope of the linear least squares fit to the 3.6-8.0~$\mu$m SEDs in log~$\lambda F_{\lambda}$~vs~log~$\lambda$ space, defined as $\alpha_{IRAC}$, was used to separate sources by evolutionary class.
In the case of the two sources we have classified as Class~III but are classified as Class~II in \citet{hart05}, the $\alpha_{IRAC}$ values are -3.07 and -2.66, both consistent with photospheric SEDs according to \citet{lada06}.
Similarly, the two sources we call Class~II that are considered Class~I in \citet{hart05} have $\alpha_{IRAC}$ values of -1.59 and -1.32, consistent with Class~II sources \citep{lada06}. 
Regardless of the ambiguity of the classifications for these few sources, this comparison demonstrates the overall efficacy of the new classification technique.




\subsection{Additional YSOs Identified via a $JHK_S[3.6][4.5]$ YSO Classification Scheme}

NGC~1333 is a moderately dense cluster with significant structured nebulosity (scattered light, $H_2$ shocks, and PAH emission), and both high stellar density and bright, diffuse emission can reduce local source sensitivity.
This issue is made worse by the decreasing angular resolution, large and bright PSF wings, and intrinsically lower sensitivity of the longer wavelength mosaics from the 5.8 and 8.0~$\mu$m IRAC channels and MIPS 24~$\mu$m. 
As mentioned above, it has been shown that $[3.6]-[4.5]$ is a useful YSO evolutionary class discriminant color \citep{alle04,mege04,hart05}, but suffers from degeneracy with differential extinction.
\citet{gt05} utilized dereddened 2-8~$\mu$m SED slopes to demonstrate that one could separate protostars, stars with disks, and diskless photospheres quite effectively with the $K-[3.6]$ vs. $[3.6]-[4.5]$ color-color diagram, provided that the colors were dereddened via the reddening law of \citet{flah07}.  
This reddening law was adopted because it was consistently measured through several nearby, star-forming molecular clouds, and shown to be significantly different than the reddening law from diffuse ISM dust \citep{inde05}. 
This empirical method of YSO identification from \citet{gt05} is explained below. 

First we measure the line of sight extinction to each source as represented by the $\frac{E_{J-H}}{E_{H-K}}$ color excess ratio, utilizing baseline colors based on the Classical T~Tauri Star (CTTS) locus of \citet{mch97} and standard dwarf star colors \citep{bb98}.  
To accomplish the latter, we force $[J-H]_0 \ge 0.6$, a simplifying approximation for the intrinsic colors of low mass dwarfs.
Here are the equations used to derive the adopted intrinsic colors from the photometry we have measured: 
\begin{displaymath}
[J-H]_0 = 0.58 \times [H-K]_{0} + 0.52
\end{displaymath}
\begin{displaymath}
[H-K]_0 = [H-K]_{meas} - ([J-H]_{meas} - [J-H]_{0}) \times \frac{E_{H-K}}{E_{J-H}}
\end{displaymath}
\begin{displaymath}
[H-K]_{0} = \frac{[J-H]_{meas} - \frac{E_{J-H}}{E_{H-K}} \times [H-K]_{meas} - 0.52}{0.58-\frac{E_{J-H}}{E_{H-K}}}
\end{displaymath}

Once we have measured the component of the $H-K$ color excess that is caused by reddening, we compute the dereddened $K-[3.6]$ and $[3.6]-[4.5]$ colors using the color excess ratios presented in \citet{flah07}, specifically $\frac{E_{J-H}}{E_{H-K}}=1.73$, $\frac{E_{H-K}}{E_{K-[3.6]}}=1.49$, and $\frac{E_{H-K}}{E_{K-[4.5]}}=1.17$: 
\begin{displaymath}
[K-[3.6]]_0 = [K-[3.6]]_{meas} - ([H-K]_{meas} - [H-K]_0) \times \frac{E_{K-[3.6]}}{E_{H-K}}
\end{displaymath}
\begin{displaymath}
[[3.6]-[4.5]]_0 = [[3.6]-[4.5]]_{meas} - ([H-K]_{meas} - [H-K]_0) \times \frac{E_{[3.6]-[4.5]}}{E_{H-K}}
\end{displaymath}
\begin{displaymath}
\frac{E_{[3.6]-[4.5]}}{E_{H-K}} = \left\{ \left[ \frac{E_{H-K}}{E_{K-[4.5]}} \right]^{-1} - \left[ \frac{E_{H-K}}{E_{K-[3.6]}} \right]^{-1} \right\}^{-1}
\end{displaymath}

We now identify additional YSOs as those sources with dereddened colors that obey all of the following constraints:
\begin{displaymath}
[[3.6]-[4.5]]_0 > 0.101
\end{displaymath}
\begin{displaymath}
[K-[3.6]]_0 > 0
\end{displaymath}
\begin{displaymath}
[K-[3.6]]_0 > -2.85714 \times ([[3.6]-[4.5]]_0 - 0.101) + 0.5
\end{displaymath}

Among these YSOs, sources that also follow the following selection are considered likely protostars (the rest are considered Class~II):
\begin{displaymath}
[K-[3.6]]_0 > -2.85714 \times ([[3.6]-[4.5]]_0 - 0.401) + 1.7  
\end{displaymath}

To minimize the inclusion of dim extragalactic contaminants, we apply a simple brightness limit using the dereddened 3.6~$\mu$m photometry.  
All sources classified as Class~II with this method must have $[3.6]_0 < 14.5$ and all protostars must have $[3.6]_0 < 15$.
In total, the use of this supplemental method allowed us to consider 504 more sources, and we added one Class~I and four Class~II sources to the NGC~1333 YSO list.

\subsection{Adding and Checking YSOs with MIPS 24~$\mu$m Photometry\label{mipsphase}}

We utilize 24~$\mu$m photometry from the MIPS instrument in several ways.  
First, sources that were considered photospheric in the previous two phases but have excess emission at 24~$\mu$m ($[5.8] - [24] > 1.5$) are very interesting.  
These sources are a mixture of regenerative debris disks and ``Transition Disks''; the latter are Class~II sources with significant dust clearing within their inner disks. 
We have required that valid sources in this group also have $[3.6] < 14$ in an effort to minimize contamination from extragalactic contaminants.

Second, the NGC~1333 cluster is well known partly because there are many known Class~0 protostellar members driving spectacular outflows. 
Many of these sources are so deeply embedded that we cannot extract usable photometry in one or more of the four IRAC bands.  
As such, we note any source as a likely deeply embedded protostar\footnote{Classifying sources as bona fide Class 0 protostars requires a detailed treatment including submillimeter flux measurements which are beyond the scope of this paper. As such, these sources are still labelled Class I in Table~\ref{ysotable2}, but with the addition of an asterisk to note their anomalous nature.} if it lacks detections in some IRAC bands yet has bright MIPS 24~$\mu$m photometry ($[24] < 7$ and $[X]-[24] > 4.5$ mags, where $[X]$ is the photometry for any IRAC detection that we possess).
The adopted magnitude limit at 24~$\mu$m is needed to mitigate extragalactic source contamination.
The specific value that we adopt here is slightly more conservative than that adopted by c2d \citep{harv06}. 
With the addition of the 24~$\micron$ data, we have more information to filter lower luminosity YSOs from our AGN candidates and shock emission dominated sources flagged in Section~\ref{iracclass}.
We reinclude flagged sources as likely protostars if they have both bright MIPS 24~$\mu$m photometry ($[24] < 7$, as before) and convincingly red IRAC/MIPS colors ($[3.6] - [5.8] > 0.5$ and $[4.5]-[24] > 4.5$ and $[8.0]-[24]  > 4$).

Finally, all previously identified protostars that have 24~$\mu$m detections are checked to ensure that their SEDs do indeed continue to rise from IRAC to MIPS wavelengths.
All protostars that have MIPS detections must have $[5.8] - [24] > 4$ if they possess 5.8~$\mu$m photometry, otherwise they must have $[4.5] - [24] > 4$.
If a source does not meet this color requirement, it is likely a highly reddened Class~II.
Of the 29 sources identified as protostars above, we find only one source that was mistakenly classified as a protostar.
We reclassify this source as a Class~II accordingly.


In total, we identify four transition/debris disks using the above integration of MIPS 24~$\mu$m data.
Of the 99 sources with MIPS detections and incomplete IRAC photometry, we add 11 deeply embedded sources to our list of protostars.
Many of the latter deeply embedded sources are known protostars from the literature, such as IRAS 4a, 4b, and 4c.
We also reclassified one deeply reddened Class~II that was first classified as Class~I in Section~\ref{iracclass} above.

\subsection{YSO Census Summary}

\begin{figure}
\epsscale{1}
\plotone{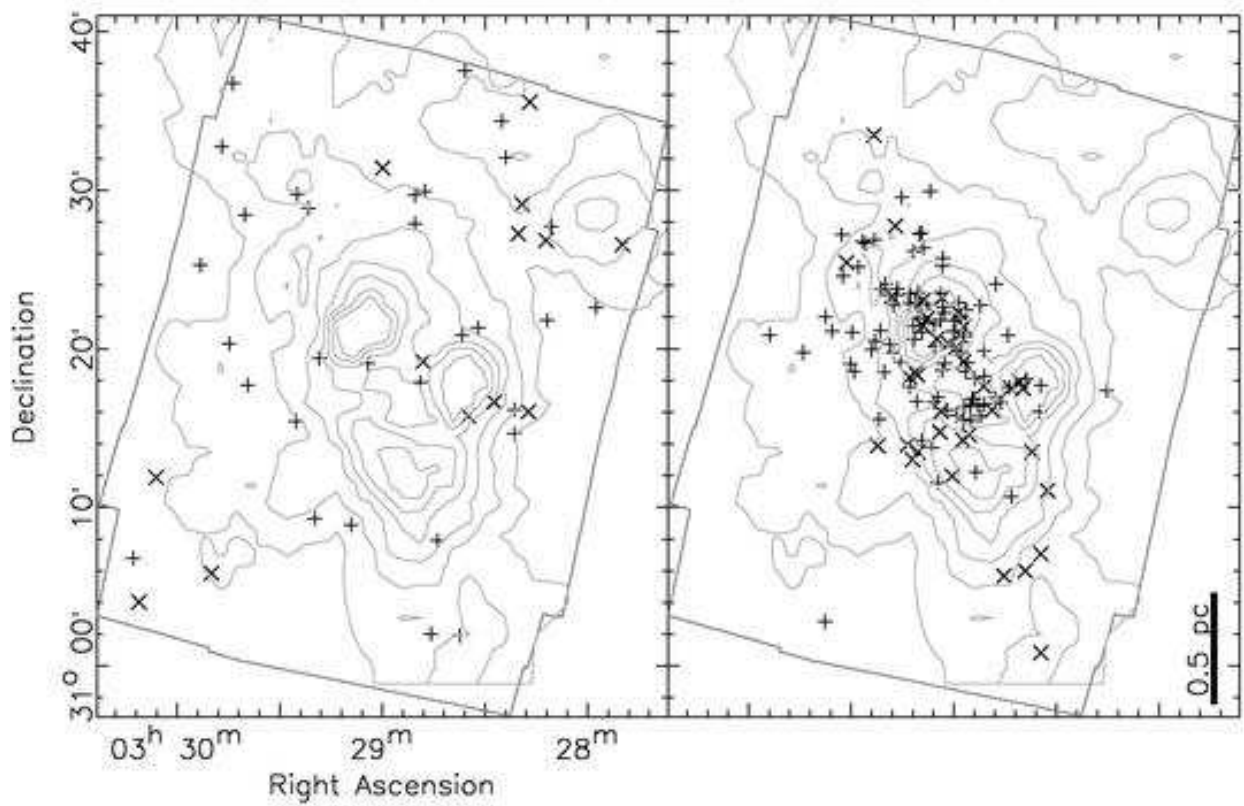}
\caption{Spatial distribution of IR excess sources in the NGC 1333 region.  At left are those sources flagged as likely extragalactic contaminants.  Crosses mark PAH-emission sources and plusses mark likely broad-line AGN.  Note that their distribution is relatively uniform and has little to do with the cluster at the center of the field of view, nor the extinction distribution overlaid in gray contours ($A_V = 3,5,7,9,11,13,15$).  At right are the identified YSO sources, overplotted on the same extinction contours.  Crosses mark all Class~I sources and plusses mark the more-evolved Class~II sources.  The clustered distribution of these sources and their close association with the extinction distribution are distinct.\label{ysomap}}
\end{figure}

\begin{figure}
\epsscale{.7}
\plotone{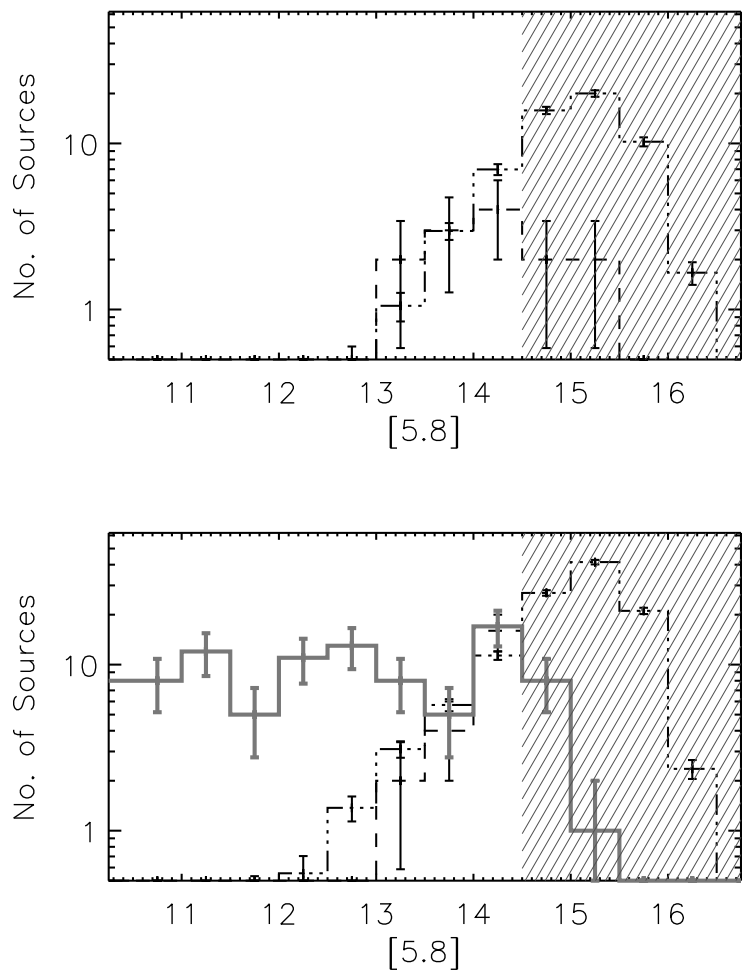}
\caption{$[5.8]$ magnitude histogram comparing numbers of sources filtered as contaminants compared to the model prediction derived from the Bootes field analysis presented in Appendix~\ref{bootes}.  The top histogram is for the PAH emission sources, and the bottom is for the likely broad-line AGN.  Bootes field models, scaled for the NGC~1333 coverage area and adjusted for relative areas of differing extinction, are plotted as black dot-dashed lines.  The distribution of sources filtered from the NGC~1333 source list are plotted as dashed lines.  The distribution of all IR excess sources in NGC~1333, excluding the PAH emission sources, is plotted as the gray solid line.  The parallel slanted line shading marks the region of the plot for sources dimmer than the 90\% completeness limit of $[5.8] = 14.5$.\label{vermhist}}
\end{figure}

For the entire three phase method, we find 4 transition/debris disks, 94 Class~II sources, and 39 protostars, for a total membership with disks of 137 sources in NGC~1333. 
For comparison to other studies, see the top plot of Figure~\ref{ctts1} for the $[3.6]-[4.5]$~vs.~$[5.8]-[8.0]$ diagram used in several recent {\it Spitzer}-based papers to identify and/or characterize YSOs \citep[e.g.][]{alle04,mege04,hart05}.
The photometry and classifications are presented in Tables~\ref{ysotable}~\&~\ref{ysotable2} (Note that the entries and indexing of these tables are in ascending Right Ascension order {/it separated by YSO class} as documented in Table~\ref{ysotable2}.).
For the rest of this paper, we consider the transition/debris disks as part of the Class~II group, yielding a total of 98 Class~II sources for the analyses that follow.
The spatial distribution of these sources is shown in the right plot of Figure~\ref{ysomap}.
While the sources are indeed quite concentrated compared to the extragalactic contaminants in the left plot of Fig.~\ref{ysomap}, the structure of the NGC~1333 cluster is visibly non-uniform.

\section{Spatial Distributions of YSOs by Class}

With a confident census of the YSOs in this region, we now examine the spatial distribution of these sources in relation to each other and that of the dense natal gas.
In Figure \ref{scubadistro}a, we have plotted the distribution of known pre-main sequence stars, namely the Class~II and transition/debris disk sources, on contours of 850~$\mu$m dust continuum emission.  
This population of sources reflects the ``double cluster'' distribution reported by \citet{lal96}.  
The two main local peaks in stellar density are confined to local minima in the dust emission.
In contrast, we have plotted the distribution of likely protostars on the same contours in Figure \ref{scubadistro}b.  
Not surprisingly, these sources trace the dense gas more closely, and thus they bear little resemblance to the ``double cluster'' distribution.  

Overall, the two-dimensional morphology of the combined star-gas system is strongly filamentary, and bears the appearance of a central elongated ring with three filaments extending away from the ring radially, roughly evenly spaced in azimuth.
The nodes where these filaments meet the central ring structure seem occupied by many young stars, while the center of the ring has a noticable absence of stars and gas.
A ring-like morphology in dense gas has been detected before in the Coalsack \citep{lada04}, and some models of collapsing clouds have yielded various ring-like morphologies \citep[e.g.][]{ln02,bh04}.
Thus, it is possible that the appearance of ring-like structure in NGC~1333 is not a spurious effect of multiple filamentary structures partly overlapping along the same line of sight, but this possibility cannot be ruled out with these observations. 
 
\subsection{Source Spacing and Proximity to Dense Gas\label{distro1}}

Here, we analyze the spatial distributions of the separate YSO populations and how they relate to the dense gas in a quantitative way. 
Figure~\ref{1v2cdf} is a histogram of the SCUBA 850~$\mu$m fluxes at each source's position for the Class~I and Class~II sources separately.
As demonstrated qualitatively in Figure~\ref{scubadistro}, protostellar sources are much more strongly confined to the regions of dense gas traced by the submillimeter emission than their more evolved Class~II counterparts.
From these data, we choose a flux threshold of 0.05 Jy/beam to demonstrate that most (27/39 = 69.2\%) Class~I sources trace dense gas, while a considerably smaller fraction of Class~II sources (20/98 = 20.4\%) are found in regions of such high submillimeter flux.


However, while their relationship to the dense gas is quite different, Figure~\ref{nnd_hist} shows that the nearest neighbor spacings between sources are quite similar for each evolutionary class, as well as for the entire set of YSOs combined.  
There is a clear peak in each histogram between 0.024~pc (5000~AU) and 0.048~pc (10000~AU).
A similar peak has recently been found in the nearest neighbor spacings of the ``Spokes'' cluster of protostars in NGC~2264, where most protostars are separated by 12000~AU to 20000~AU \citep{teix06}.
The peak in nearest neighbor spacings among the NGC~1333 protostars and stars with disks is half of the value found for the ``Spokes'' cluster. 
However, this peak is consistent with the clump and YSO separations of $\sim$6000~AU observed in the Ophiuchus region by \citet{man98}. 


\subsection{Surface Density Mapping}

Following the method presented in \citet{gute05}, we have constructed surface density maps for the identified YSO distribution in NGC~1333 using the Nearest Neighbor method\footnote{In summary, maps are generated by measuring the area A of the smallest circle that contains the $n$ nearest stars centered at regular grid intervals; the local surface density at the $i_{th}$ grid point is simply $\sigma_i = (n-1) / A_i$ and the uncertainty in each measurement is equal to the density divided by $\sqrt{n-2}$ \citep{ch85}.
We will be referencing nearest neigbor maps of differing $n$ throughout this work, thus we adopt the notation NN$n$ to represent this concept.}. 
Figure~\ref{nngs} shows a comparison between maps generated using $n$~=~4,~6,~11,~and~18 neighbors.
The smallest $n$ value maps reveal structure on very small scales.  
Most of the surface density structure in these high resolution maps traces the dense gas structure in the cloud quite closely.
However, as we increase $n$ and approach measurements expected to be more robust in the face of Poisson statistics (for example, NN18 implies a surface density signal to noise ratio of 4 on any given measurement \citep{ch85}), most structure disappears, leaving behind only the two-peaked nature of the ``double cluster''.


\section{Fraction of Members with Disks\label{df}}

It has been demonstrated previously that the fraction of cluster members with disks roughly traces the mean age of the pre-main sequence cluster members \citep[e.g.][]{hll01,hern07}.
Here we present a measurement of this quantity for NGC~1333 in order to place it within the context of this trend.

Based on the overall structure observed in the density maps, we have adopted a circular region labeled ``Main'' in Figure~\ref{dfregion} that is clearly dominated by cluster members for measurement of the fraction of members that have disks.  
The circle is centered on the median Right Ascension and Declination of the {\it Spitzer}-identified YSOs, and has a radius of 5.5 arcminutes, or 0.4~pc.  
We have adopted a rectangular field star reference region, also labeled in Figure~\ref{dfregion}, where no YSOs are found and little structure is present in the extinction map or the SCUBA map.  
Finally, we have chosen several smaller regions of the cluster for similar analysis to examine if the disk fraction changes between the northern and southern peaks in stellar density, as well as the diffuse Class~II-dominated region to the east of the main cluster. 

Figure \ref{dfhist} demonstrates the $K_S$ magnitude histogram (KMH) analysis that yields an estimate of the total membership, including diskless pre-main sequence stars, within the ``Main'' region down to the adopted magnitude limit of $K_S = 14$ for all sources that have at least $H$, $K_S$, $[3.6]$, and $[4.5]$ photometry.
A detailed description of the application of this process can be found in Section~3 of \citet{gute05}.  
In summary, we count the number of stars per 0.5 magnitude bin in the cluster region and in the reference region and take the difference, applying a correction to the reference region distribution to account for the mean difference in extinction and the ratio of the areas between the two regions.
After field star contamination is accounted for, we estimate that 87 members are located within the ``Main'' cluster region.  
The adopted $K_S$ magnitude limit, corresponding to a 1~Myr old, 0.08~$M_{\odot}$ pre-main sequence star \citep{bara98} behind $A_K = 2$~mag of extinction, was adopted to ensure uniform detection likelihood in sufficient near-IR and mid-IR wavelengths to enable confident classification of sources.
Further, this cut, in combination with the $H$-band completeness limit of 16.1, achieves an approximate mass- and extinction--limited sampling of cluster members \citep{carp97}.  
In this way, we effectively mitigate biases against detecting diskless members over members with disks (the longer wavelength {\it Spitzer} photometry is deeper that the near-IR photometry, enabling detection of both types of sources for a given near-IR magnitude cut) and detecting low-luminosity members {\it with disks} that may have been cut in the elimination of extragalactic contaminants (though the number of these sources appears to be small, see Fig.~\ref{vermhist} in Section~\ref{iracclass}).  
The total number of {\it Spitzer}-identified YSOs with $K_S < 14$ in the same region is 72, yielding a disk fraction of $83\% \pm 11\%$.

We have summarized the disk fraction measurements for all subregions in Table~\ref{dftable}.
All measurements are consistent to within the uncertainty of the field star contamination estimates, suggesting that both components of the ``double cluster'' as well as the more diffuse population to the east are roughly coeval.
For further comparison, \citet{wilk04} estimated a disk fraction for the northern subcluster of $75\% \pm 20\%$, consistent with our results, using their K-band excess fraction with a correction factor applied to account for this method's incomplete census of stars with disks. 

\section{1D Surface Density Profiles}

Most recent models of clustered star formation do not attempt to predict specific radial density profile falloff behavior, as such models suggest that high dynamical activity dominates these environments, making any structure in a simulated cluster quite transient.  
However, recent $N_2H^+$ measurements of the radial velocities of cold cores in NGC~1333 show that the velocity dispersion is very low \citep{wals07}.
An analysis of the distances of red MIPS sources in NGC~1333 and the rest of Perseus from submillimeter cores suggests that the protostellar sources have a very low velocity dispersion relative to the dense gas \citep{jorg06b}.
Furthermore, the similarity in the nearest neighbor spacings between the Class~I and Class~II sources is also indicative that the relative speeds of the cluster members is small and the Class~II sources are particularly young.
Therefore, the structure of NGC~1333 is particularly intriguing as it should offer a direct probe of an early phase of the clustered star formation process, before gravitational interactions have had a chance to erase structure in the member distribution. 
Here we measure the radial density profile for NGC~1333, as well as measure the degree of asymmetry in the cluster.  

\subsection{Radial Density Profile: Annular Star Counts Method\label{radstd1}}

Unlike the regions examined in \cite{gute05}, the density of cluster members relative to the field star density in the NGC~1333 region is too low to allow for any meaningful measurement of the radial profile of the cluster from the $K_S$-band source list.
Thus, we restrict this analysis of the character of the cluster radial profile to the one measured for the {\it Spitzer}-identified YSOs only.  Given the large fraction of the total membership inferred to have disks, this should introduce minimal additional uncertainty to the analysis. 

Here we adopt the method of measuring the radial surface density profile described in detail in \cite{gute05}.  
In summary, once a cluster center point is adopted, sources are counted in concentric annular regions of constant radial extent, and the counts are divided by the area of each annulus.  
The bin size used here is 1~arcminute, or roughly 0.07~pc. 
However, choosing the center point is an uncertain task for clusters that are non-uniform and/or asymmetric.

Given the degree of substructure in NGC~1333, we have measured radial profiles from four distinct center points: ``Main'' is the median of the Right Ascension and Declination of the YSOs; ``Ring Center'' is the visual center of the cluster, located at the center of the apparent ring traced by the dense gas and YSOs; ``North'' and ``South'' are the locations of peak density in the two main maxima of the double cluster. 
See Figure~\ref{radprof1} for the resulting profiles. 
The YSOs in NGC~1333 are distributed roughly uniformly in the inner 0.3~pc radius region of the cluster.  
The choice of center point primarily affects the short radius behavior of the profile, depending on proximity to a local maximum.
Also, the profiles made from offset center points (``North'' and ``South'') have clear contamination from the other local maximum at $\sim$0.55~pc.
This ambiguity demonstrates that traditional radial profile techniques like this are not particularly well-suited for studies of young clusters.


\citet{gute05} fit measured radial profiles with exponential decay functions, but noted without demonstration that the functional choice seemed to be arbitrary; one could fit any number of well-behaved functions to the cluster radial profiles measured for those clusters given the uncertainty introduced by field star contamination.
The radial profiles derived from the NGC~1333 YSOs do not remotely resemble an exponential, nor even a single index power law, however.
The dashed lines in Figure~\ref{radprof1} are two power laws chosen to emulate the functional behavior of the ``Main'' and ``Ring Center'' profiles in the inner and outer radius regimes.  
Inside the knee at 0.3~pc, the profile is approximately flat (power law index of $\alpha=0$).
Outside that knee, the descent is precipitous, corresponding to a power law index of $\alpha=-3.3$.
Alternatively, one can also model the behavior of the profile with a slightly more complicated function, such as the red curve in Figure~\ref{radprof1}.
This is a function of the form $\sigma = \sigma_0 (1+(\frac{r}{r_0})^\beta)^{-\gamma/\beta}$, with parameters $\sigma_0 = 220$~pc$^{-2}$, $r_0 = 0.3$~pc, $\beta = 6$, and $\gamma = 3.3$.

The function above is very similar in functional form to the EFF profile often used to characterize the radial profiles of young massive clusters in other galaxies \citep{eff87}, although the value of $\beta$ in that work is kept constant at $\beta=2$.
We allowed that value to vary, yielding a closer fit to the entire profile.
For relatively extreme fitting parameters ( $\sigma_0 = 220$~pc$^{-2}$, $r_0 = 0.7$~pc, and $\gamma = 8.0$), an EFF profile can be fit to these data.
Similarly, we can fit a King profile \citep{king62} of central density $k = 400$~pc$^{-2}$, core radius $r_{c} = 0.4$~pc, and truncation radius $r_t = 1.3$~pc.
See the right column of plots in Figure~\ref{radprof1} for overlays of these functions on the measured radial profiles.
Regardless of the fitting results, the physical inferences one can normally derive from these models are not valid, as NGC~1333 is a very non-uniform, asymmetric system (see Section~\ref{aap} below) that appears to be dominated by structure in the dense gas.

\subsection{Azimuthal Distribution Histogram\label{aap}}

We next apply another one dimensional analysis of structure, averaging radially rather than azimuthally in order to measure the azimuthal distribution histogram \citep[ADH; ][]{gute05} for the YSOs in NGC~1333.
The ADH is presented in Figure~\ref{azhist1}.  Non-uniform azimuthal structure is apparent, with peaks at PA$\sim30^{\circ}$ and $\sim210^{\circ}$.
These peaks correspond to the ``double cluster'' local density maxima.   
To quantify the asymmetry we see here, we measure the {\it azimuthal asymmetry parameter} (AAP) of this ADH, defined as the ratio of the standard deviation of the histogram to the square root of its mean value in \citet{gute05}.
The AAP is meant to quantify the degree of asymmetry measured in an ADH with respect to the deviations expected from a normal, circularly symmetric cluster distribution, e.g. one characterized by Poisson variability only.  
Monte Carlo simulations presented in \citet{gute05} demonstrated that for clusters of $N\sim100$ sources, the behavior of the AAP of an unstructured cluster is reasonably approximated as a Gaussian with a mean value of $\sim$0.9 and a standard deviation of $\sim$0.2.
Thus the AAP of 2 measured here for NGC~1333 is quite high, demonstrating that the cluster is significantly deviant from a circularly symmetric distribution.

\section{Discussion}

We have employed the mid-IR sensitivity of {\it Spitzer} to achieve a census of the YSO members of the NGC~1333 embedded cluster that was previously unattainable.
Furthermore, we have shown that the sources identified in our {\it Spitzer} census represent a large fraction of the total cluster membership (83\%). 
With this penetrating view of the cluster, we have performed several measurements of different aspects of the structure of the cluster that are not confused by field star contamination and are less biased by extinction effects than previous studies.

We confirm the double-peaked surface density morphology of the cluster reported in previous work \citep{lal96}, with the caveat that this morphology is traced only by the more evolved Class~II population.
The two main density peaks are located in local minima of the gas density distribution.
In contrast, the protostars trace the dense gas in this region closely, as do a fraction of the Class~II.
That gas distribution appears to be a network of filamentary structures, connecting the two density peaks into a single, albeit complex, structure.
Despite the difference in spatial distributions, however, the Class~II and protostars have similarities in their nearest neighbor distance distributions, particularly in the location of the peak at spacings of 0.045~pc.
These two evolutionary states are expected to differ in duration by an order of magnitude, thus a low overall velocity dispersion in the cluster and a very young Class~II population are needed to account for the similarity remaining in the spacings among the two populations.


A further implication of the lack of evidence for dynamical evolution of the cluster is the need for the dispersal of the molecular gas by the Class~II sources on very short timescales.
Given that the more evolved YSOs have not moved significantly from their birthsites, regions dominated by more evolved sources where gas densities are preferentially low, such as the double-peaks in the stellar distribution here, were once locations of high gas density.
Thus it seems plausible that groups of low-mass stars are able to disrupt their dense natal gas {\it locally}. 
In this sense, we concur with previous work that has suggested that the active outflows in NGC~1333 are destroying the cloud \citep{sk01,quil05}.  
With a virtual lack of low or high mass stars and dense gas in the center of the ring-like structure of the cloud and cluster core, we argue that this is in fact a primordial structure and not outflow-evacuated.
 
Considering all sources regardless of evolutionary class, the cluster is clearly elongated, an expected result given the double-peaked nature of the stellar surface density distribution and embedded nature of the cluster as a whole \citep{gute05}. 
We have presented radial density profiles measured via several different methods, and all suggest a roughly uniform density distribution within a 0.3~pc radius, with a steep decline ($\alpha=-3.3$) beyond. 
The flat central distribution and sharp radial decline in the surface density profile at radii larger than 0.3~pc also suggests a rather limited amount of dynamical interaction in this cluster.
If we analyze this within the framework of dynamics-generated structures like those of King or EFF models, we have to choose extreme fitting parameters just to get marginal agreement with the measured profiles. 
The radial profile we have measured here has two clear regimes were a simple power law matchs the density profile well, and the sharp knee transition between them is poorly matched by either the King or EFF profiles.
Furthermore, given the asymmetry of the NGC~1333 cluster and the presense of a considerable amount of structured, dense gas, the dynamical information inferred from either of these models is unlikely to be accurate.
Cloud geometry is the most tenable cause for the structure we have observed.
As such, the relative motions of the stars must be fairly slow and the Class~II population must be quite young in order to have preserved that structure into the Class~II evolutionary phase.  

\section{Acknowledgements}

We wish to thank T. Huard for numerous discussions on extragalactic contamination characterization and mitigation.  
We also thank R. Cutri for providing the deeper 2MASS observations ahead of public release.
This publication makes use of data products from the Two Micron All Sky Survey, which is a joint project of the University of Massachusetts and the Infrared Processing and Analysis Center/California Institute of Technology, funded by the National Aeronautics and Space Administration and the National Science Foundation.
This research has made use of the SIMBAD database, operated at CDS, Strasbourg, France.
This work is based in part on observations made with the Spitzer Space Telescope, which is operated by the Jet Propulsion Laboratory, California Institute of Technology under a contract with NASA.






\appendix

\section{Extragalactic Contamination Characterization and Mitigation\label{bootes}}

While IR-excess emission is a potent membership diagnostic for young and embedded clusters, there are many potential contaminants.  
For NGC~1333, and most clusters in the {\it Spitzer} Young Stellar Clusters Survey, the dominant source of contamination is from extragalactic sources\footnote{For regions closer to the galactic plane or nearer to the galactic center, evolved stars are also relatively bright and numerous, but this source of contamination is not expected to be significant for most nearby star-forming regions.  Thus, we do not addressed that in this paper.}.
\citet{ster05} demonstrated that there were two classes of extragalactic sources that have some form of IR-excess emission and were quite numerous once sufficiently deep integrations are obtained with IRAC.  
First, normal star-forming galaxies and narrow-line AGN have growing excess at 5.8 and 8.0~$\mu$m, as PAH emission excited by young, high-mass stars dominates their largely stellar SEDs in these bandpasses.  
Second, broad-line AGN have red, non-stellar SEDs, resulting in IRAC colors that are very similar to those of bona fide YSOs.  

To aid in the development of mitigation strategies for the {\it Spitzer} Young Cluster survey and studies of nearby star formation in general, we have obtained the Bootes field IRAC data used in \citet{ster05} from the {\it Spitzer} Science Center online public archive and reduced $\sim$7.7 square degrees of it using identical methods to those presented in Section~\ref{obs}.
The total exposure time in the mosaicked data is twice that of the NGC~1333 mosaics, and the field lacks significant structured nebulosity. 
Thus the Bootes data are marginally deeper than NGC~1333, moreso in the 5.8 and 8.0~$\mu$m mosaics than the 3.6 and 4.5~$\mu$m ones.
Without any attempt at cutting out contaminants, our IRAC four channel color-based classification scheme would suggest that we had detected 1268 Class~I and 2641 Class~II YSOs in this field; clearly a false result. 
\citet{harv06} and \citet{jorg06a} suggested a basic cut developed for the c2d Legacy Survey to filter these sources: cut all sources with $[4.5] < 14$.
This method is relatively effective for the Bootes data in that only 80 (10.4 deg$^{-2}$) Class~I and 108 (14 deg$^{-2}$) Class~II of the above numbers of IR-excess sources would bypass the filter.  

In exploring the efficacy of the c2d filtering method, we determined a few simple refinements that seem to improve the overall performance of the method. 
We present the modified method here, along with a characterization of the number density of sources per magnitude for verification of this and other extragalactic filtering methods.
In Figures~\ref{bpah1}-\ref{bhist1}, we show select color-color and color-magnitude diagrams to give a sense of the regions of these diagrams where these non-YSO IR-excess sources fall, their approximate number densities as a function of survey depth, and a suggested set of color and magnitude cuts that can filter out most of these sources while minimizing loss of bona fide YSOs.

First, we exploit the result from \cite{ster05} that galaxies dominated by PAH emission have mid-IR colors that occupy a relatively unique area of most IRAC color-color diagrams.
One simply has to filter out those sources that fall within either of the following regions of two IRAC color-color spaces. 
For the $[4.5]-[5.8]$~vs.~$[5.8]-[8.0]$ diagram (see the top plot in Figure~\ref{bpah1}), we reject sources that follow all these conditions:
\begin{displaymath}
[4.5]-[5.8] < \frac{1.05}{1.2}\times([5.8]-[8.0]-1)
\end{displaymath}
\begin{displaymath}
[4.5]-[5.8] < 1.05
\end{displaymath}
\begin{displaymath}
[5.8]-[8.0] > 1
\end{displaymath}

For the $[3.6]-[5.8]$~vs.~$[4.5]-[8.0]$ diagram (see bottom plot in Figure~\ref{bpah1}), we reject sources that follow all these conditions:
\begin{displaymath}
[3.6]-[5.8] < \frac{1.5}{2}\times([4.5]-[8.0]-1)
\end{displaymath}
\begin{displaymath}
[3.6]-[5.8] < 1.5
\end{displaymath}
\begin{displaymath}
[4.5]-[8.0] > 1
\end{displaymath}

Only a small number of true YSOs have been shown to have significant PAH emission in other regions of young and forming groups of stars \citep{hart05,hern07}.
Since no PAH emission source in the Bootes field has $[4.5]<11.5$, we require all sources removed to be dimmer than this value.
This ensures that at least the PAH-rich YSOs known at this time will not be excluded using this cut.

The broad-line AGN have mid-IR colors that are largely consistent with YSOs \citep{ster05}.
Thus, to filter these sources from IRAC source lists for nearby star-forming regions, one must exploit the fact that the apparent magnitudes of the extragalactic sources are much dimmer than those of most YSOs within the nearest kiloparsec.
By filtering out the PAH sources first, we can construct a filter that more closely traces the broad-line AGN distribution than the original c2d proposed cut. 
We have developed this modified cut after applying considerable effort to characterizing the contaminants from the Bootes field and finding ways to isolate them from YSOs.
Following the c2d lead \citep{harv06}, we utilize the $[4.5]$~vs.~$[4.5]-[8.0]$ color-magnitude diagram to flag likely AGN.
However, with the PAH sources removed separately, we have refined the AGN cutoff, flagging all sources that follow all of the following three conditions:

\begin{displaymath}
[4.5]-[8.0] > 0.5
\end{displaymath}
\begin{displaymath}
[4.5] > 13.5+([4.5]-[8.0]-2.3)/0.4
\end{displaymath}
\begin{displaymath}
[4.5]>13.5
\end{displaymath}

Additionally, a source flagged as a likely AGN must follow any one of the following three conditions:

\begin{displaymath}
[4.5] > 14+([4.5]-[8.0]-0.5)
\end{displaymath}
\begin{displaymath}
[4.5] > 14.5-([4.5]-[8.0]-1.2)/0.3
\end{displaymath}
\begin{displaymath}
[4.5] > 14.5
\end{displaymath}

In Figure~\ref{bagn1}, we demonstrate the results of the above cutoff for the Bootes data.  
This cut has been chosen to avoid cutting large sections of color or magnitude space where we expect YSOs in the nearest kiloparsec to fall, thus this method does allow some residual contamination.
However, this set of cuts is a significant improvement over the c2d filtration method.
When applied to the Bootes data, it flags 1485 sources as PAH emission sources and 2766 as likely AGN, leaving behind 49 (6.4~deg$^{-2}$) false Class 0/I sources and 29 (3.8~deg$^{-2}$) false Class~II sources as residual contamination.
 
Since most of these contaminating sources (PAH and AGN) are relatively bright at 8.0~$\mu$m, it appears that the band that limits sensitivity to these contaminant sources is the 5.8~$\mu$m channel.  
Therefore, it is useful to characterize the $[5.8]$ magnitude distribution of each class of source from the Bootes dataset to demonstrate the effectiveness of the method presented here and model the expected residual contamination in a statistical sense (see Figure~\ref{bhist1}).
The differential surface density distributions in units of number of sources per square degree per half-magnitude bin are both represented reasonably well by the following power laws:

\begin{displaymath}
\sigma_{PAH} = 10^{-18.24} \times [5.8]^{17.02} 
\end{displaymath}
\begin{displaymath}
\sigma_{BL-AGN} = 10^{-20.90} \times [5.8]^{19.63} 
\end{displaymath}

Also, while differential completeness limits are not very accurately determined using magnitude histograms such as these, the overall completeness ``decay'' behavior can be measured from the PAH emission source distribution, since these sources are uniquely identifiable.  
This behavior can then be applied to the power-law model above for the more ambiguously identified AGN distribution in order to help to compute the total number of these sources down to the detection limit of a given set of photometry data.

\clearpage



\begin{deluxetable}{cccccc}
\tabletypesize{\scriptsize}
\tablecaption{Summary of Photometric Data Parameters\label{photpar}}
\tablewidth{0pt}
\tablehead{
\colhead{Bandpass} & \colhead{Flux for Zero Mag.(Jy)} & \colhead{Mag. for 90\% Completeness\tablenotemark{a}} & \colhead{Median Unc. at Completeness Limit} & \colhead{Mag. for 1 DN/s\tablenotemark{b}}  &  \colhead{$\frac{A_{\lambda}}{A_K}$\tablenotemark{c}}
}
\startdata
$J$     & 1669  & 16.8 & 0.06 & ...    & 2.50 \\
$H$     & 980   & 16.1 & 0.08 & ...    & 1.55 \\
$K_S$   & 620   & 15.3 & 0.06 & ...    & 1.00 \\
$[3.6]$ & 280.9 & 17.0 & 0.08 & 19.455 & 0.63 \\
$[4.5]$ & 179.7 & 16.5 & 0.09 & 18.699 & 0.53 \\
$[5.8]$ & 115.0 & 14.5 & 0.11 & 16.498 & 0.49 \\
$[8.0]$ & 64.13 & 13.8 & 0.12 & 16.892 & 0.49 \\
$[24]$  & 7.3   &  9.4 & 0.08 & 14.600 & 0.43 \\
\enddata
\tablenotetext{a}{2MASS $JHK_S$ values inferred from 2MASS Online Documentation; {\it Spitzer} values computed using method detailed in \citet{gute05}.}
\tablenotetext{b}{Photometric calibration values account for adopted apertures.}
\tablenotetext{c}{Adopted from \citet{flah07}.}
\end{deluxetable}

\begin{deluxetable}{ccccccccccc}
\tabletypesize{\scriptsize}
\rotate
\tablecaption{{\it Spitzer}-identified YSOs: 2MASS, IRAC, and MIPS Photometry\label{ysotable}}
\tablewidth{0pt}
\tablehead{\colhead{Index} & \colhead{R.A.\tablenotemark{a}} & \colhead{Decl.\tablenotemark{a}} & \colhead{$J$} & \colhead{$H$} & \colhead{$K_S$} & \colhead{$[3.6]$} & \colhead{$[4.5]$} & \colhead{$[5.8]$} & \colhead{$[8.0]$} & \colhead{$[24]$} }
\startdata
1 & 03:28:39.09 & +31:06:01.9 & ... $\pm$ ... & ... $\pm$ ... & ... $\pm$ ... & $15.62 \pm 0.04$ & $14.26 \pm 0.03$ & $13.31 \pm 0.05$ & $11.91 \pm 0.03$ & $6.17 \pm 0.01$ \\
2 & 03:28:45.30 & +31:05:42.1 & ... $\pm$ ... & $16.75 \pm 0.15$ & $14.91 \pm 0.06$ & $13.17 \pm 0.02$ & $11.82 \pm 0.01$ & $11.37 \pm 0.02$ & $10.72 \pm 0.01$ & $3.73 \pm 0.01$ \\
3 & 03:28:57.37 & +31:14:16.2 & ... $\pm$ ... & ... $\pm$ ... & ... $\pm$ ... & ... $\pm$ ... & $7.81 \pm 0.01$ & $6.60 \pm 0.01$ & $5.61 \pm 0.01$ & $0.47 \pm 0.01$ \\
4 & 03:29:00.50 & +31:12:00.6 & ... $\pm$ ... & ... $\pm$ ... & ... $\pm$ ... & ... $\pm$ ... & $15.11 \pm 0.09$ & ... $\pm$ ... & ... $\pm$ ... & $6.82 \pm 0.03$ \\
5 & 03:29:04.06 & +31:14:46.7 & ... $\pm$ ... & ... $\pm$ ... & ... $\pm$ ... & $14.21 \pm 0.04$ & $12.81 \pm 0.04$ & ... $\pm$ ... & $11.54 \pm 0.03$ & $5.17 \pm 0.06$ \\
6 & 03:29:10.49 & +31:13:30.8 & ... $\pm$ ... & ... $\pm$ ... & ... $\pm$ ... & ... $\pm$ ... & $15.48 \pm 0.07$ & ... $\pm$ ... & ... $\pm$ ... & $6.13 \pm 0.04$ \\
7 & 03:29:11.25 & +31:18:31.7 & ... $\pm$ ... & ... $\pm$ ... & ... $\pm$ ... & $13.95 \pm 0.13$ & $11.61 \pm 0.06$ & $10.81 \pm 0.06$ & $10.44 \pm 0.06$ & $2.68 \pm 0.02$ \\
8 & 03:29:12.04 & +31:13:02.0 & ... $\pm$ ... & ... $\pm$ ... & ... $\pm$ ... & $13.95 \pm 0.01$ & $10.47 \pm 0.01$ & $10.95 \pm 0.01$ & $10.70 \pm 0.04$ & $5.55 \pm 0.02$ \\
9 & 03:29:13.60 & +31:13:58.2 & ... $\pm$ ... & ... $\pm$ ... & ... $\pm$ ... & $15.69 \pm 0.04$ & $13.53 \pm 0.02$ & $12.83 \pm 0.06$ & $13.12 \pm 0.18$ & $6.00 \pm 0.03$ \\
10 & 03:29:17.16 & +31:27:46.1 & ... $\pm$ ... & ... $\pm$ ... & ... $\pm$ ... & $14.28 \pm 0.02$ & $12.78 \pm 0.01$ & $12.25 \pm 0.03$ & $12.16 \pm 0.05$ & $4.52 \pm 0.01$ \\
11 & 03:29:23.48 & +31:33:29.4 & ... $\pm$ ... & ... $\pm$ ... & ... $\pm$ ... & ... $\pm$ ... & $12.35 \pm 0.01$ & $11.58 \pm 0.02$ & $10.90 \pm 0.02$ & $3.75 \pm 0.01$ \\
12 & 03:28:32.55 & +31:11:04.8 & ... $\pm$ ... & ... $\pm$ ... & $14.49 \pm 0.12$ & $13.22 \pm 0.02$ & $12.31 \pm 0.01$ & $11.56 \pm 0.01$ & $10.64 \pm 0.01$ & $5.29 \pm 0.01$ \\
13 & 03:28:34.49 & +31:00:51.2 & $14.72 \pm 0.03$ & $13.09 \pm 0.04$ & $11.89 \pm 0.03$ & $9.12 \pm 0.01$ & $8.20 \pm 0.01$ & $6.81 \pm 0.01$ & $6.20 \pm 0.01$ & $1.89 \pm 0.01$ \\
14 & 03:28:34.52 & +31:07:05.5 & ... $\pm$ ... & ... $\pm$ ... & ... $\pm$ ... & $12.15 \pm 0.01$ & $10.55 \pm 0.01$ & $9.37 \pm 0.01$ & $8.25 \pm 0.01$ & $5.18 \pm 0.01$ \\
15 & 03:28:37.09 & +31:13:31.1 & ... $\pm$ ... & ... $\pm$ ... & $12.05 \pm 0.03$ & $9.81 \pm 0.01$ & $8.01 \pm 0.01$ & $6.67 \pm 0.01$ & $4.95 \pm 0.01$ & ... $\pm$ ... \\
16 & 03:28:39.69 & +31:17:32.0 & ... $\pm$ ... & $17.04 \pm 0.20$ & $13.81 \pm 0.03$ & $10.81 \pm 0.01$ & $9.69 \pm 0.01$ & $8.88 \pm 0.01$ & $8.00 \pm 0.01$ & $4.03 \pm 0.01$ \\
17 & 03:28:40.62 & +31:17:56.5 & ... $\pm$ ... & ... $\pm$ ... & ... $\pm$ ... & $16.93 \pm 0.18$ & $14.93 \pm 0.05$ & $13.44 \pm 0.05$ & $12.30 \pm 0.05$ & $7.71 \pm 0.13$ \\
18 & 03:28:43.27 & +31:17:33.1 & $12.48 \pm 0.02$ & $10.84 \pm 0.05$ & $9.51 \pm 0.04$ & $7.82 \pm 0.01$ & $6.86 \pm 0.01$ & $5.91 \pm 0.01$ & $4.90 \pm 0.01$ & $1.32 \pm 0.01$ \\
19 & 03:28:48.76 & +31:16:08.9 & ... $\pm$ ... & $16.84 \pm 0.16$ & $14.95 \pm 0.05$ & $13.19 \pm 0.01$ & $12.39 \pm 0.01$ & $11.48 \pm 0.01$ & $10.42 \pm 0.01$ & $6.50 \pm 0.02$ \\
20 & 03:28:51.27 & +31:17:39.5 & $16.02 \pm 0.05$ & $14.83 \pm 0.08$ & $14.31 \pm 0.06$ & $13.91 \pm 0.03$ & $12.96 \pm 0.02$ & $12.09 \pm 0.02$ & $10.49 \pm 0.01$ & $6.02 \pm 0.02$ \\
21 & 03:28:55.52 & +31:14:35.8 & ... $\pm$ ... & ... $\pm$ ... & ... $\pm$ ... & $12.66 \pm 0.01$ & $10.18 \pm 0.01$ & $8.96 \pm 0.01$ & $8.33 \pm 0.01$ & $0.35 \pm 0.01$ \\
22 & 03:28:57.09 & +31:21:25.1 & ... $\pm$ ... & ... $\pm$ ... & $16.37 \pm 0.14$ & $14.35 \pm 0.03$ & $13.59 \pm 0.03$ & $12.87 \pm 0.04$ & $12.12 \pm 0.03$ & ... $\pm$ ... \\
23 & 03:28:57.11 & +31:19:11.9 & $17.21 \pm 0.08$ & $15.97 \pm 0.09$ & $15.24 \pm 0.05$ & $14.03 \pm 0.04$ & $13.08 \pm 0.04$ & $12.22 \pm 0.05$ & $11.12 \pm 0.05$ & $7.26 \pm 0.10$ \\
24 & 03:28:57.99 & +31:20:52.1 & ... $\pm$ ... & ... $\pm$ ... & ... $\pm$ ... & $15.47 \pm 0.08$ & $14.37 \pm 0.07$ & $13.09 \pm 0.07$ & $11.89 \pm 0.04$ & ... $\pm$ ... \\
25 & 03:28:58.42 & +31:22:17.7 & $17.99 \pm 0.16$ & $14.43 \pm 0.04$ & $11.56 \pm 0.03$ & $9.60 \pm 0.01$ & $8.01 \pm 0.01$ & $7.04 \pm 0.01$ & $5.96 \pm 0.01$ & $2.28 \pm 0.01$ \\
26 & 03:28:59.31 & +31:20:08.2 & $17.80 \pm 0.14$ & ... $\pm$ ... & $15.10 \pm 0.05$ & $14.21 \pm 0.09$ & $13.25 \pm 0.07$ & $12.23 \pm 0.16$ & $11.18 \pm 0.09$ & ... $\pm$ ... \\
27 & 03:29:01.54 & +31:20:20.7 & ... $\pm$ ... & $13.88 \pm 0.08$ & $10.88 \pm 0.04$ & $6.92 \pm 0.01$ & $5.73 \pm 0.01$ & $4.53 \pm 0.01$ & $3.50 \pm 0.01$ & ... $\pm$ ... \\
28 & 03:29:03.32 & +31:23:14.8 & $17.63 \pm 0.11$ & $15.48 \pm 0.06$ & $14.02 \pm 0.04$ & $12.00 \pm 0.01$ & $10.98 \pm 0.01$ & $9.49 \pm 0.01$ & $7.93 \pm 0.01$ & $4.31 \pm 0.02$ \\
29 & 03:29:03.76 & +31:16:04.0 & $11.74 \pm 0.02$ & $9.74 \pm 0.05$ & $8.15 \pm 0.02$ & $6.22 \pm 0.01$ & $5.02 \pm 0.01$ & $3.92 \pm 0.01$ & $2.73 \pm 0.01$ & ... $\pm$ ... \\
30 & 03:29:04.94 & +31:20:38.6 & ... $\pm$ ... & $14.22 \pm 0.07$ & $12.65 \pm 0.04$ & $11.30 \pm 0.01$ & $10.52 \pm 0.01$ & $9.72 \pm 0.01$ & $8.78 \pm 0.01$ & $4.80 \pm 0.04$ \\
31 & 03:29:07.77 & +31:21:57.3 & ... $\pm$ ... & $13.80 \pm 0.09$ & $10.43 \pm 0.04$ & $6.65 \pm 0.01$ & $5.13 \pm 0.01$ & $4.13 \pm 0.01$ & $3.65 \pm 0.03$ & ... $\pm$ ... \\
32 & 03:29:08.95 & +31:22:56.3 & $16.45 \pm 0.07$ & $13.76 \pm 0.05$ & $11.78 \pm 0.03$ & $9.98 \pm 0.02$ & $9.05 \pm 0.01$ & $8.63 \pm 0.11$ & ... $\pm$ ... & ... $\pm$ ... \\
33 & 03:29:09.07 & +31:21:29.1 & ... $\pm$ ... & $16.10 \pm 0.10$ & $13.18 \pm 0.03$ & $10.30 \pm 0.01$ & $8.39 \pm 0.01$ & $7.42 \pm 0.01$ & $6.74 \pm 0.03$ & $2.34 \pm 0.05$ \\
34 & 03:29:09.09 & +31:23:05.7 & $14.61 \pm 0.03$ & $12.90 \pm 0.03$ & $11.84 \pm 0.03$ & $10.65 \pm 0.02$ & $10.10 \pm 0.02$ & $8.97 \pm 0.10$ & $7.33 \pm 0.14$ & $2.83 \pm 0.12$ \\
35 & 03:29:10.71 & +31:18:21.2 & ... $\pm$ ... & ... $\pm$ ... & ... $\pm$ ... & $12.57 \pm 0.01$ & $10.82 \pm 0.01$ & $9.77 \pm 0.01$ & $9.18 \pm 0.01$ & $2.27 \pm 0.02$ \\
36 & 03:29:12.95 & +31:18:14.6 & ... $\pm$ ... & ... $\pm$ ... & $14.14 \pm 0.03$ & $9.28 \pm 0.01$ & $7.53 \pm 0.01$ & $6.64 \pm 0.01$ & $5.80 \pm 0.01$ & $2.60 \pm 0.01$ \\
37 & 03:29:18.25 & +31:23:19.9 & ... $\pm$ ... & ... $\pm$ ... & ... $\pm$ ... & $13.53 \pm 0.05$ & $11.99 \pm 0.02$ & $10.63 \pm 0.02$ & $9.59 \pm 0.03$ & $4.17 \pm 0.02$ \\
38 & 03:29:22.28 & +31:13:54.5 & ... $\pm$ ... & ... $\pm$ ... & ... $\pm$ ... & $14.26 \pm 0.01$ & $13.29 \pm 0.01$ & $12.50 \pm 0.02$ & $11.14 \pm 0.01$ & $7.73 \pm 0.03$ \\
39 & 03:29:31.54 & +31:25:27.8 & $17.18 \pm 0.07$ & $16.03 \pm 0.08$ & $15.35 \pm 0.06$ & $14.34 \pm 0.01$ & $13.56 \pm 0.01$ & $12.78 \pm 0.03$ & $12.17 \pm 0.06$ & $7.90 \pm 0.09$ \\
40 & 03:28:15.19 & +31:17:23.8 & $16.51 \pm 0.05$ & $15.75 \pm 0.07$ & $15.13 \pm 0.05$ & $14.45 \pm 0.01$ & $13.98 \pm 0.01$ & $13.49 \pm 0.05$ & $12.82 \pm 0.05$ & ... $\pm$ ... \\
41 & 03:28:34.47 & +31:17:43.2 & ... $\pm$ ... & $17.13 \pm 0.19$ & $15.20 \pm 0.05$ & $13.96 \pm 0.01$ & $13.39 \pm 0.01$ & $12.96 \pm 0.03$ & $12.14 \pm 0.03$ & $8.50 \pm 0.05$ \\
42 & 03:28:34.85 & +31:16:04.5 & $17.55 \pm 0.10$ & $16.37 \pm 0.10$ & $15.13 \pm 0.05$ & $13.69 \pm 0.01$ & $13.10 \pm 0.01$ & $12.48 \pm 0.02$ & $11.61 \pm 0.02$ & $8.03 \pm 0.05$ \\
43 & 03:28:38.76 & +31:18:06.7 & ... $\pm$ ... & $16.83 \pm 0.15$ & $14.16 \pm 0.03$ & $12.00 \pm 0.01$ & $10.97 \pm 0.01$ & $10.41 \pm 0.01$ & $9.56 \pm 0.01$ & $6.82 \pm 0.04$ \\
44 & 03:28:43.23 & +31:10:42.7 & $15.29 \pm 0.03$ & $13.15 \pm 0.03$ & $12.04 \pm 0.02$ & $11.20 \pm 0.01$ & $10.88 \pm 0.01$ & $10.57 \pm 0.01$ & $10.23 \pm 0.01$ & $8.13 \pm 0.04$ \\
45 & 03:28:43.56 & +31:17:36.5 & $12.02 \pm 0.03$ & $10.79 \pm 0.03$ & $9.85 \pm 0.05$ & $8.90 \pm 0.01$ & $8.38 \pm 0.01$ & $7.77 \pm 0.01$ & $6.64 \pm 0.01$ & ... $\pm$ ... \\
46 & 03:28:44.08 & +31:20:52.9 & $14.21 \pm 0.02$ & $13.21 \pm 0.03$ & $12.61 \pm 0.03$ & $11.90 \pm 0.01$ & $11.68 \pm 0.01$ & $11.27 \pm 0.01$ & $10.61 \pm 0.01$ & $8.07 \pm 0.03$ \\
47 & 03:28:46.19 & +31:16:38.7 & $10.88 \pm 0.02$ & $10.01 \pm 0.02$ & $9.69 \pm 0.02$ & $9.47 \pm 0.01$ & $9.17 \pm 0.01$ & $8.88 \pm 0.01$ & $8.30 \pm 0.01$ & ... $\pm$ ... \\
48 & 03:28:47.63 & +31:24:06.1 & $14.26 \pm 0.02$ & $12.61 \pm 0.03$ & $11.64 \pm 0.02$ & $10.34 \pm 0.01$ & $9.97 \pm 0.01$ & $9.66 \pm 0.01$ & $9.27 \pm 0.01$ & $6.52 \pm 0.01$ \\
49 & 03:28:47.82 & +31:16:55.3 & $12.93 \pm 0.02$ & $11.75 \pm 0.03$ & $10.92 \pm 0.02$ & $9.94 \pm 0.01$ & $9.33 \pm 0.01$ & $9.01 \pm 0.01$ & $8.33 \pm 0.01$ & $5.60 \pm 0.02$ \\
50 & 03:28:51.02 & +31:18:18.5 & $11.36 \pm 0.02$ & $10.07 \pm 0.02$ & $9.18 \pm 0.02$ & $8.50 \pm 0.01$ & $7.84 \pm 0.01$ & $7.55 \pm 0.01$ & $6.49 \pm 0.01$ & $3.37 \pm 0.01$ \\
51 & 03:28:51.07 & +31:16:32.6 & $13.29 \pm 0.02$ & $12.50 \pm 0.03$ & $12.14 \pm 0.02$ & $11.91 \pm 0.01$ & $11.41 \pm 0.01$ & $11.07 \pm 0.01$ & $10.37 \pm 0.01$ & $8.29 \pm 0.07$ \\
52 & 03:28:51.19 & +31:19:54.9 & $11.72 \pm 0.02$ & $10.49 \pm 0.02$ & $9.90 \pm 0.02$ & $9.08 \pm 0.01$ & $8.65 \pm 0.01$ & $8.36 \pm 0.01$ & $7.29 \pm 0.01$ & $4.31 \pm 0.01$ \\
53 & 03:28:52.13 & +31:15:47.2 & $13.19 \pm 0.02$ & $12.50 \pm 0.03$ & $12.07 \pm 0.02$ & $11.48 \pm 0.01$ & $11.21 \pm 0.01$ & $10.77 \pm 0.01$ & $10.04 \pm 0.01$ & $7.79 \pm 0.07$ \\
54 & 03:28:52.15 & +31:22:45.4 & $11.98 \pm 0.02$ & $11.01 \pm 0.02$ & $10.56 \pm 0.02$ & $10.18 \pm 0.01$ & $9.94 \pm 0.01$ & $9.56 \pm 0.01$ & $8.92 \pm 0.01$ & $6.32 \pm 0.02$ \\
55 & 03:28:52.90 & +31:16:26.6 & $13.59 \pm 0.02$ & $12.87 \pm 0.03$ & $12.50 \pm 0.02$ & $12.26 \pm 0.01$ & $11.85 \pm 0.01$ & $11.42 \pm 0.01$ & $10.82 \pm 0.02$ & $8.29 \pm 0.17$ \\
56 & 03:28:53.58 & +31:12:14.7 & ... $\pm$ ... & $16.80 \pm 0.15$ & $15.38 \pm 0.06$ & $14.10 \pm 0.01$ & $13.58 \pm 0.01$ & $13.09 \pm 0.04$ & $12.34 \pm 0.04$ & $9.07 \pm 0.12$ \\
57 & 03:28:53.93 & +31:18:09.3 & $14.57 \pm 0.03$ & $12.22 \pm 0.03$ & $10.92 \pm 0.03$ & $10.09 \pm 0.01$ & $9.69 \pm 0.01$ & $9.15 \pm 0.01$ & $8.30 \pm 0.01$ & $4.82 \pm 0.01$ \\
58 & 03:28:54.07 & +31:16:54.5 & $13.02 \pm 0.02$ & $12.05 \pm 0.05$ & $11.61 \pm 0.02$ & $11.19 \pm 0.01$ & $10.65 \pm 0.01$ & $10.21 \pm 0.01$ & $9.31 \pm 0.01$ & $5.12 \pm 0.01$ \\
59 & 03:28:54.61 & +31:16:51.3 & $12.82 \pm 0.02$ & $11.22 \pm 0.03$ & $10.32 \pm 0.02$ & $9.54 \pm 0.01$ & $8.79 \pm 0.01$ & $8.45 \pm 0.01$ & $7.61 \pm 0.01$ & $4.50 \pm 0.02$ \\
60 & 03:28:54.92 & +31:15:29.2 & $16.02 \pm 0.04$ & $14.97 \pm 0.05$ & $14.18 \pm 0.03$ & $13.55 \pm 0.01$ & $13.07 \pm 0.01$ & $12.65 \pm 0.04$ & $11.99 \pm 0.04$ & ... $\pm$ ... \\
61 & 03:28:55.07 & +31:16:28.8 & $13.46 \pm 0.02$ & $11.61 \pm 0.03$ & $10.51 \pm 0.04$ & $9.23 \pm 0.01$ & $8.47 \pm 0.01$ & $8.14 \pm 0.01$ & $7.46 \pm 0.01$ & $3.91 \pm 0.02$ \\
62 & 03:28:55.15 & +31:16:24.8 & $12.96 \pm 0.03$ & $11.59 \pm 0.05$ & $10.81 \pm 0.03$ & $9.93 \pm 0.01$ & $9.50 \pm 0.01$ & $9.08 \pm 0.01$ & $8.30 \pm 0.01$ & ... $\pm$ ... \\
63 & 03:28:56.09 & +31:19:08.6 & ... $\pm$ ... & ... $\pm$ ... & $16.18 \pm 0.12$ & $11.22 \pm 0.01$ & $9.71 \pm 0.01$ & $8.84 \pm 0.01$ & $8.26 \pm 0.01$ & $6.14 \pm 0.04$ \\
64 & 03:28:56.31 & +31:22:28.0 & $15.53 \pm 0.03$ & $13.07 \pm 0.03$ & $11.77 \pm 0.03$ & $10.61 \pm 0.01$ & $10.00 \pm 0.01$ & $9.27 \pm 0.01$ & $8.36 \pm 0.01$ & $4.75 \pm 0.03$ \\
65 & 03:28:56.64 & +31:18:35.7 & $12.30 \pm 0.02$ & $10.69 \pm 0.02$ & $9.70 \pm 0.02$ & $8.83 \pm 0.01$ & $8.35 \pm 0.01$ & $8.25 \pm 0.01$ & $7.55 \pm 0.01$ & $3.67 \pm 0.01$ \\
66 & 03:28:56.93 & +31:20:48.6 & $15.47 \pm 0.03$ & $14.48 \pm 0.04$ & $13.87 \pm 0.03$ & $13.31 \pm 0.01$ & $13.17 \pm 0.01$ & $12.77 \pm 0.03$ & $11.89 \pm 0.04$ & ... $\pm$ ... \\
67 & 03:28:56.95 & +31:16:22.3 & $13.72 \pm 0.02$ & $11.89 \pm 0.03$ & $11.03 \pm 0.02$ & $10.61 \pm 0.01$ & $10.31 \pm 0.01$ & $9.92 \pm 0.01$ & $8.93 \pm 0.01$ & $3.97 \pm 0.02$ \\
68 & 03:28:57.17 & +31:15:34.6 & $15.38 \pm 0.03$ & $14.00 \pm 0.03$ & $13.20 \pm 0.03$ & $12.61 \pm 0.01$ & $11.90 \pm 0.01$ & $11.26 \pm 0.02$ & $10.27 \pm 0.01$ & $7.17 \pm 0.09$ \\
69 & 03:28:57.70 & +31:19:48.1 & $13.05 \pm 0.02$ & $11.95 \pm 0.03$ & $11.38 \pm 0.03$ & $10.62 \pm 0.01$ & $10.10 \pm 0.01$ & $9.71 \pm 0.01$ & $8.67 \pm 0.01$ & $5.76 \pm 0.05$ \\
70 & 03:28:58.24 & +31:22:09.4 & $16.04 \pm 0.04$ & $14.32 \pm 0.04$ & $13.32 \pm 0.03$ & $12.46 \pm 0.02$ & $11.96 \pm 0.03$ & $11.36 \pm 0.07$ & ... $\pm$ ... & ... $\pm$ ... \\
71 & 03:28:58.25 & +31:22:02.1 & $14.55 \pm 0.02$ & $13.09 \pm 0.03$ & $12.22 \pm 0.02$ & $11.30 \pm 0.01$ & $10.84 \pm 0.01$ & $10.30 \pm 0.01$ & $9.59 \pm 0.04$ & ... $\pm$ ... \\
72 & 03:28:58.43 & +31:22:56.8 & $15.31 \pm 0.03$ & $14.24 \pm 0.04$ & $13.65 \pm 0.03$ & $13.00 \pm 0.01$ & $12.73 \pm 0.01$ & $12.58 \pm 0.05$ & $12.07 \pm 0.04$ & ... $\pm$ ... \\
73 & 03:28:59.32 & +31:15:48.5 & $16.16 \pm 0.04$ & $12.54 \pm 0.03$ & $10.56 \pm 0.02$ & $8.69 \pm 0.01$ & $8.01 \pm 0.01$ & $7.50 \pm 0.01$ & $6.40 \pm 0.01$ & $2.46 \pm 0.01$ \\
74 & 03:28:59.56 & +31:21:46.8 & $12.37 \pm 0.02$ & $10.96 \pm 0.03$ & $9.98 \pm 0.05$ & $8.85 \pm 0.01$ & $8.40 \pm 0.01$ & $8.14 \pm 0.01$ & $7.51 \pm 0.01$ & $4.77 \pm 0.03$ \\
75 & 03:29:00.15 & +31:21:09.3 & $16.61 \pm 0.05$ & $14.60 \pm 0.04$ & $13.35 \pm 0.03$ & $12.11 \pm 0.01$ & $11.68 \pm 0.01$ & $11.41 \pm 0.02$ & $10.62 \pm 0.03$ & $6.46 \pm 0.12$ \\
76 & 03:29:02.16 & +31:16:11.4 & $14.51 \pm 0.02$ & $13.87 \pm 0.03$ & $13.62 \pm 0.03$ & $13.33 \pm 0.04$ & $12.76 \pm 0.06$ & ... $\pm$ ... & ... $\pm$ ... & ... $\pm$ ... \\
77 & 03:29:02.79 & +31:22:17.4 & $17.07 \pm 0.07$ & $14.88 \pm 0.04$ & $13.54 \pm 0.03$ & $12.20 \pm 0.01$ & $11.56 \pm 0.01$ & $11.07 \pm 0.02$ & $10.46 \pm 0.03$ & ... $\pm$ ... \\
78 & 03:29:03.13 & +31:22:38.2 & $14.05 \pm 0.02$ & $12.78 \pm 0.03$ & $11.73 \pm 0.02$ & $10.25 \pm 0.01$ & $9.60 \pm 0.01$ & $9.21 \pm 0.01$ & $8.23 \pm 0.01$ & $5.00 \pm 0.07$ \\
79 & 03:29:03.20 & +31:25:45.3 & $15.66 \pm 0.03$ & $14.62 \pm 0.04$ & $13.98 \pm 0.03$ & $13.22 \pm 0.01$ & $12.64 \pm 0.01$ & $12.24 \pm 0.03$ & $11.56 \pm 0.05$ & $7.34 \pm 0.09$ \\
80 & 03:29:03.39 & +31:18:40.1 & $15.86 \pm 0.03$ & $14.65 \pm 0.04$ & $13.97 \pm 0.03$ & $13.26 \pm 0.01$ & $13.02 \pm 0.02$ & $12.68 \pm 0.04$ & $12.37 \pm 0.04$ & ... $\pm$ ... \\
81 & 03:29:03.42 & +31:25:14.5 & $16.32 \pm 0.05$ & $14.87 \pm 0.05$ & $13.93 \pm 0.04$ & $13.04 \pm 0.01$ & $12.67 \pm 0.01$ & $12.15 \pm 0.02$ & $11.50 \pm 0.04$ & ... $\pm$ ... \\
82 & 03:29:03.86 & +31:21:48.8 & $11.43 \pm 0.02$ & $10.16 \pm 0.03$ & $9.36 \pm 0.04$ & $8.05 \pm 0.01$ & $7.62 \pm 0.01$ & $7.24 \pm 0.01$ & $6.32 \pm 0.01$ & $3.50 \pm 0.03$ \\
83 & 03:29:03.93 & +31:23:31.2 & $17.20 \pm 0.08$ & $16.06 \pm 0.07$ & $15.17 \pm 0.06$ & $13.95 \pm 0.01$ & $13.45 \pm 0.01$ & $12.90 \pm 0.07$ & $12.42 \pm 0.17$ & ... $\pm$ ... \\
84 & 03:29:04.67 & +31:16:59.2 & $15.61 \pm 0.03$ & $13.95 \pm 0.03$ & $12.75 \pm 0.02$ & $11.40 \pm 0.01$ & $10.85 \pm 0.01$ & $10.24 \pm 0.01$ & $9.22 \pm 0.01$ & $5.64 \pm 0.14$ \\
85 & 03:29:04.73 & +31:11:35.0 & ... $\pm$ ... & ... $\pm$ ... & $14.31 \pm 0.03$ & $12.04 \pm 0.01$ & $11.03 \pm 0.01$ & $10.41 \pm 0.01$ & $9.82 \pm 0.01$ & $7.39 \pm 0.04$ \\
86 & 03:29:05.64 & +31:20:10.7 & $17.20 \pm 0.07$ & $16.34 \pm 0.09$ & $15.65 \pm 0.08$ & $13.98 \pm 0.02$ & $13.39 \pm 0.02$ & $12.48 \pm 0.02$ & $11.75 \pm 0.03$ & ... $\pm$ ... \\
87 & 03:29:05.67 & +31:21:33.8 & ... $\pm$ ... & $15.25 \pm 0.05$ & $13.31 \pm 0.03$ & $11.70 \pm 0.01$ & $10.92 \pm 0.01$ & $10.33 \pm 0.02$ & $9.75 \pm 0.05$ & ... $\pm$ ... \\
88 & 03:29:05.76 & +31:16:39.7 & $14.29 \pm 0.02$ & $11.58 \pm 0.03$ & $9.82 \pm 0.06$ & $8.56 \pm 0.01$ & $7.82 \pm 0.01$ & $7.16 \pm 0.01$ & $6.32 \pm 0.01$ & $3.71 \pm 0.05$ \\
89 & 03:29:06.32 & +31:13:46.5 & ... $\pm$ ... & $14.93 \pm 0.04$ & $12.75 \pm 0.02$ & $11.06 \pm 0.01$ & $10.29 \pm 0.01$ & $9.70 \pm 0.01$ & $9.00 \pm 0.01$ & $5.97 \pm 0.03$ \\
90 & 03:29:06.92 & +31:29:57.1 & $16.70 \pm 0.05$ & $15.55 \pm 0.05$ & $14.83 \pm 0.04$ & $14.06 \pm 0.01$ & $13.79 \pm 0.01$ & $13.48 \pm 0.04$ & $12.93 \pm 0.05$ & $9.76 \pm 0.11$ \\
91 & 03:29:07.94 & +31:22:51.6 & $12.98 \pm 0.02$ & $11.17 \pm 0.03$ & $10.13 \pm 0.04$ & $8.93 \pm 0.01$ & $8.39 \pm 0.01$ & $7.95 \pm 0.01$ & $7.38 \pm 0.05$ & ... $\pm$ ... \\
92 & 03:29:08.95 & +31:26:24.1 & $17.14 \pm 0.07$ & $15.99 \pm 0.07$ & $15.73 \pm 0.08$ & $13.38 \pm 0.01$ & $12.65 \pm 0.01$ & $12.07 \pm 0.02$ & $11.26 \pm 0.03$ & $7.34 \pm 0.10$ \\
93 & 03:29:09.33 & +31:21:04.2 & $16.38 \pm 0.04$ & $14.34 \pm 0.03$ & $13.19 \pm 0.03$ & $12.28 \pm 0.01$ & $11.92 \pm 0.01$ & $11.36 \pm 0.07$ & $10.20 \pm 0.07$ & ... $\pm$ ... \\
94 & 03:29:09.47 & +31:27:21.0 & $14.11 \pm 0.02$ & $13.18 \pm 0.03$ & $12.67 \pm 0.03$ & $12.06 \pm 0.01$ & $11.71 \pm 0.01$ & $11.34 \pm 0.01$ & $10.41 \pm 0.01$ & $7.82 \pm 0.12$ \\
95 & 03:29:09.63 & +31:22:56.5 & $11.26 \pm 0.02$ & $10.14 \pm 0.05$ & $9.50 \pm 0.03$ & $9.13 \pm 0.01$ & $8.92 \pm 0.01$ & $8.33 \pm 0.08$ & $7.19 \pm 0.19$ & ... $\pm$ ... \\
96 & 03:29:10.18 & +31:27:15.9 & $15.51 \pm 0.03$ & $14.69 \pm 0.04$ & $14.20 \pm 0.03$ & $13.74 \pm 0.01$ & $13.54 \pm 0.01$ & $13.50 \pm 0.06$ & ... $\pm$ ... & ... $\pm$ ... \\
97 & 03:29:10.40 & +31:21:59.3 & $9.34 \pm 0.02$ & $7.95 \pm 0.04$ & $7.17 \pm 0.02$ & $6.70 \pm 0.01$ & $6.54 \pm 0.01$ & $5.95 \pm 0.09$ & $4.63 \pm 0.14$ & ... $\pm$ ... \\
98 & 03:29:10.46 & +31:23:35.0 & $15.65 \pm 0.03$ & $13.85 \pm 0.03$ & $12.85 \pm 0.03$ & $11.87 \pm 0.01$ & $11.41 \pm 0.01$ & $10.90 \pm 0.03$ & $10.30 \pm 0.10$ & ... $\pm$ ... \\
99 & 03:29:10.82 & +31:16:42.7 & $15.97 \pm 0.03$ & $14.47 \pm 0.04$ & $13.20 \pm 0.02$ & $11.60 \pm 0.01$ & $10.90 \pm 0.01$ & $10.16 \pm 0.01$ & $9.05 \pm 0.01$ & $5.87 \pm 0.03$ \\
100 & 03:29:11.33 & +31:22:57.1 & $14.84 \pm 0.02$ & $13.46 \pm 0.03$ & $12.58 \pm 0.02$ & $11.50 \pm 0.01$ & $11.04 \pm 0.01$ & $10.72 \pm 0.12$ & ... $\pm$ ... & ... $\pm$ ... \\
101 & 03:29:11.64 & +31:20:37.7 & $15.34 \pm 0.03$ & $13.60 \pm 0.03$ & $12.69 \pm 0.03$ & $11.93 \pm 0.01$ & $11.60 \pm 0.01$ & $11.16 \pm 0.02$ & $10.53 \pm 0.03$ & $6.02 \pm 0.08$ \\
102 & 03:29:11.77 & +31:26:09.7 & $16.87 \pm 0.06$ & $15.77 \pm 0.06$ & $14.76 \pm 0.04$ & $13.85 \pm 0.01$ & $13.29 \pm 0.01$ & $12.64 \pm 0.07$ & $11.22 \pm 0.08$ & $6.35 \pm 0.12$ \\
103 & 03:29:11.87 & +31:21:27.1 & ... $\pm$ ... & $15.52 \pm 0.05$ & $12.69 \pm 0.03$ & $10.58 \pm 0.01$ & $9.69 \pm 0.01$ & $9.00 \pm 0.03$ & $8.25 \pm 0.06$ & ... $\pm$ ... \\
104 & 03:29:12.90 & +31:23:29.5 & $13.43 \pm 0.02$ & $12.58 \pm 0.03$ & $12.10 \pm 0.03$ & $11.29 \pm 0.01$ & $10.78 \pm 0.01$ & $10.62 \pm 0.03$ & $10.09 \pm 0.07$ & ... $\pm$ ... \\
105 & 03:29:13.04 & +31:17:38.4 & $15.22 \pm 0.02$ & $14.59 \pm 0.04$ & $14.19 \pm 0.03$ & $13.55 \pm 0.01$ & $13.10 \pm 0.01$ & $12.76 \pm 0.05$ & $12.08 \pm 0.04$ & ... $\pm$ ... \\
106 & 03:29:13.13 & +31:22:52.9 & $13.04 \pm 0.02$ & $11.24 \pm 0.03$ & $10.09 \pm 0.05$ & $8.83 \pm 0.01$ & $8.38 \pm 0.01$ & $7.95 \pm 0.01$ & $7.50 \pm 0.01$ & $5.61 \pm 0.19$ \\
107 & 03:29:15.34 & +31:29:34.6 & ... $\pm$ ... & ... $\pm$ ... & $15.83 \pm 0.09$ & $14.00 \pm 0.01$ & $13.39 \pm 0.01$ & $12.66 \pm 0.03$ & $11.90 \pm 0.03$ & $9.21 \pm 0.08$ \\
108 & 03:29:15.66 & +31:19:11.1 & ... $\pm$ ... & ... $\pm$ ... & $16.47 \pm 0.15$ & $14.99 \pm 0.02$ & $14.56 \pm 0.02$ & $14.08 \pm 0.08$ & $13.47 \pm 0.09$ & ... $\pm$ ... \\
109 & 03:29:16.59 & +31:23:49.6 & $13.17 \pm 0.02$ & $11.82 \pm 0.03$ & $11.16 \pm 0.02$ & $10.32 \pm 0.01$ & $9.73 \pm 0.01$ & $9.21 \pm 0.01$ & $8.45 \pm 0.04$ & ... $\pm$ ... \\
110 & 03:29:16.81 & +31:23:25.4 & $15.25 \pm 0.03$ & $14.10 \pm 0.03$ & $13.52 \pm 0.03$ & $12.85 \pm 0.01$ & $12.42 \pm 0.01$ & $11.79 \pm 0.02$ & $10.82 \pm 0.02$ & $6.36 \pm 0.10$ \\
111 & 03:29:17.66 & +31:22:45.2 & $9.95 \pm 0.02$ & $8.94 \pm 0.02$ & $8.35 \pm 0.02$ & $7.29 \pm 0.01$ & $6.82 \pm 0.01$ & $6.28 \pm 0.01$ & $5.54 \pm 0.01$ & $2.89 \pm 0.01$ \\
112 & 03:29:17.76 & +31:19:48.2 & $14.78 \pm 0.03$ & $13.65 \pm 0.03$ & $12.97 \pm 0.03$ & $12.24 \pm 0.01$ & $11.87 \pm 0.01$ & $11.53 \pm 0.01$ & $10.93 \pm 0.01$ & $8.09 \pm 0.05$ \\
113 & 03:29:18.65 & +31:20:17.9 & ... $\pm$ ... & ... $\pm$ ... & $14.61 \pm 0.04$ & $13.55 \pm 0.01$ & $12.89 \pm 0.01$ & $12.27 \pm 0.02$ & $11.06 \pm 0.01$ & $6.30 \pm 0.03$ \\
114 & 03:29:18.72 & +31:23:25.5 & $11.37 \pm 0.02$ & $10.38 \pm 0.09$ & $10.26 \pm 0.02$ & $9.58 \pm 0.01$ & $9.10 \pm 0.01$ & $8.35 \pm 0.01$ & $7.06 \pm 0.01$ & $3.81 \pm 0.04$ \\
115 & 03:29:20.05 & +31:24:07.6 & ... $\pm$ ... & $14.69 \pm 0.05$ & $12.16 \pm 0.03$ & $9.97 \pm 0.01$ & $8.94 \pm 0.01$ & $8.32 \pm 0.04$ & $7.32 \pm 0.08$ & $3.22 \pm 0.07$ \\
116 & 03:29:20.42 & +31:18:34.3 & $14.03 \pm 0.02$ & $11.79 \pm 0.03$ & $10.39 \pm 0.02$ & $8.57 \pm 0.01$ & $7.95 \pm 0.01$ & $7.49 \pm 0.01$ & $6.59 \pm 0.01$ & $3.08 \pm 0.01$ \\
117 & 03:29:21.30 & +31:23:46.5 & $17.48 \pm 0.10$ & $14.66 \pm 0.04$ & $13.07 \pm 0.03$ & $11.68 \pm 0.01$ & $11.19 \pm 0.01$ & $10.72 \pm 0.04$ & $10.09 \pm 0.12$ & ... $\pm$ ... \\
118 & 03:29:21.55 & +31:21:10.5 & $12.41 \pm 0.02$ & $11.72 \pm 0.03$ & $11.43 \pm 0.02$ & $11.01 \pm 0.01$ & $10.68 \pm 0.01$ & $10.32 \pm 0.01$ & $9.62 \pm 0.01$ & $6.74 \pm 0.03$ \\
119 & 03:29:21.87 & +31:15:36.4 & $11.10 \pm 0.02$ & $10.05 \pm 0.03$ & $9.45 \pm 0.03$ & $8.59 \pm 0.01$ & $8.20 \pm 0.01$ & $7.89 \pm 0.01$ & $7.41 \pm 0.01$ & $4.82 \pm 0.01$ \\
120 & 03:29:23.15 & +31:20:30.5 & $12.35 \pm 0.02$ & $11.64 \pm 0.03$ & $11.30 \pm 0.02$ & $10.51 \pm 0.01$ & $10.16 \pm 0.01$ & $9.77 \pm 0.01$ & $8.93 \pm 0.01$ & $5.92 \pm 0.01$ \\
121 & 03:29:23.23 & +31:26:53.1 & $13.52 \pm 0.02$ & $12.67 \pm 0.03$ & $12.24 \pm 0.03$ & $11.63 \pm 0.01$ & $11.09 \pm 0.01$ & $10.72 \pm 0.01$ & $10.02 \pm 0.01$ & $7.51 \pm 0.14$ \\
122 & 03:29:24.08 & +31:19:57.8 & $15.55 \pm 0.04$ & $14.25 \pm 0.04$ & $13.59 \pm 0.03$ & $12.87 \pm 0.01$ & $12.44 \pm 0.01$ & $11.88 \pm 0.01$ & $10.75 \pm 0.01$ & $5.03 \pm 0.01$ \\
123 & 03:29:25.92 & +31:26:40.1 & $11.13 \pm 0.03$ & $10.11 \pm 0.04$ & ... $\pm$ ... & $8.92 \pm 0.01$ & $8.14 \pm 0.01$ & $7.67 \pm 0.01$ & $6.32 \pm 0.01$ & $4.31 \pm 0.01$ \\
124 & 03:29:28.01 & +31:25:11.0 & $17.42 \pm 0.09$ & $15.94 \pm 0.07$ & $14.89 \pm 0.04$ & $13.87 \pm 0.01$ & $13.37 \pm 0.01$ & $12.85 \pm 0.05$ & $12.28 \pm 0.11$ & ... $\pm$ ... \\
125 & 03:29:29.79 & +31:21:02.8 & $12.66 \pm 0.02$ & $11.63 \pm 0.03$ & $11.18 \pm 0.02$ & $10.69 \pm 0.01$ & $10.43 \pm 0.01$ & $9.92 \pm 0.01$ & $8.95 \pm 0.01$ & $6.08 \pm 0.03$ \\
126 & 03:29:30.39 & +31:19:03.4 & $12.12 \pm 0.02$ & $11.37 \pm 0.03$ & $11.06 \pm 0.02$ & $10.49 \pm 0.01$ & $10.12 \pm 0.01$ & $9.69 \pm 0.01$ & $8.93 \pm 0.01$ & $6.04 \pm 0.01$ \\
127 & 03:29:32.56 & +31:24:37.0 & $14.31 \pm 0.02$ & $12.90 \pm 0.03$ & $11.85 \pm 0.02$ & $10.52 \pm 0.01$ & $10.06 \pm 0.01$ & $9.80 \pm 0.01$ & $9.26 \pm 0.01$ & $6.54 \pm 0.05$ \\
128 & 03:29:32.86 & +31:27:12.7 & $13.37 \pm 0.02$ & $12.66 \pm 0.03$ & $12.32 \pm 0.03$ & $11.76 \pm 0.01$ & $11.46 \pm 0.01$ & $11.23 \pm 0.01$ & $10.73 \pm 0.02$ & $8.51 \pm 0.10$ \\
129 & 03:29:35.70 & +31:21:08.6 & ... $\pm$ ... & ... $\pm$ ... & ... $\pm$ ... & $15.06 \pm 0.02$ & $13.96 \pm 0.01$ & $13.33 \pm 0.05$ & $12.77 \pm 0.05$ & ... $\pm$ ... \\
130 & 03:29:37.63 & +31:02:49.3 & $14.68 \pm 0.02$ & $14.13 \pm 0.03$ & $13.77 \pm 0.03$ & $13.18 \pm 0.01$ & $13.01 \pm 0.01$ & $12.48 \pm 0.03$ & $11.76 \pm 0.03$ & $9.03 \pm 0.05$ \\
131 & 03:29:37.72 & +31:22:02.6 & $13.95 \pm 0.02$ & $13.37 \pm 0.02$ & $12.97 \pm 0.02$ & $12.35 \pm 0.01$ & $12.06 \pm 0.01$ & $11.71 \pm 0.01$ & $11.00 \pm 0.01$ & $7.34 \pm 0.03$ \\
132 & 03:29:44.15 & +31:19:47.5 & $17.46 \pm 0.10$ & $16.59 \pm 0.12$ & $15.16 \pm 0.05$ & $13.38 \pm 0.01$ & $12.67 \pm 0.01$ & $12.07 \pm 0.01$ & $11.46 \pm 0.02$ & $9.02 \pm 0.06$ \\
133 & 03:29:54.03 & +31:20:53.1 & $11.93 \pm 0.02$ & $11.13 \pm 0.02$ & $10.60 \pm 0.02$ & $9.90 \pm 0.01$ & $9.50 \pm 0.01$ & $9.15 \pm 0.01$ & $8.27 \pm 0.01$ & $4.46 \pm 0.01$ \\
134 & 03:29:02.69 & +31:19:05.8 & $17.71 \pm 0.12$ & $15.31 \pm 0.05$ & $13.63 \pm 0.03$ & $12.21 \pm 0.01$ & $11.50 \pm 0.01$ & $11.30 \pm 0.01$ & $11.17 \pm 0.02$ & $7.39 \pm 0.06$ \\
135 & 03:29:09.41 & +31:14:14.1 & ... $\pm$ ... & ... $\pm$ ... & ... $\pm$ ... & $12.44 \pm 0.01$ & $10.91 \pm 0.01$ & $10.26 \pm 0.01$ & $10.16 \pm 0.04$ & $7.43 \pm 0.10$ \\
136 & 03:29:26.79 & +31:26:47.7 & $10.83 \pm 0.02$ & $9.99 \pm 0.03$ & $9.68 \pm 0.03$ & $9.58 \pm 0.01$ & $9.40 \pm 0.01$ & $9.28 \pm 0.01$ & $8.96 \pm 0.01$ & $5.91 \pm 0.02$ \\
137 & 03:29:29.26 & +31:18:34.8 & $12.58 \pm 0.02$ & $11.40 \pm 0.03$ & $10.98 \pm 0.02$ & $10.67 \pm 0.01$ & $10.65 \pm 0.01$ & $10.52 \pm 0.01$ & $10.05 \pm 0.01$ & $4.71 \pm 0.01$ \\
\enddata
\tablenotetext{a}{J2000 coordinates}
\end{deluxetable}

\begin{deluxetable}{ccccccc}
\tabletypesize{\scriptsize}
\rotate
\tablecaption{{\it Spitzer}-identified YSOs: Addendum\label{ysotable2}}
\tablewidth{0pt}
\tablehead{\colhead{Index} & \colhead{R.A.\tablenotemark{a}} & \colhead{Decl.\tablenotemark{a}} & \colhead{Lit. Matches} & \colhead{Class\tablenotemark{b}} & \colhead{$A_K$\tablenotemark{c}} & \colhead{$\alpha_{IRAC}$\tablenotemark{d}} }
\startdata
1 & 03:28:39.09 & +31:06:01.9 & ... & I* & ... & 1.35 \\
2 & 03:28:45.30 & +31:05:42.1 & IRAS 03256+3055 & I* & ... & -0.17 \\
3 & 03:28:57.37 & +31:14:16.2 & IRAS 2b & I* & ... & ... \\
4 & 03:29:00.50 & +31:12:00.6 & ... & I* & ... & ... \\
5 & 03:29:04.06 & +31:14:46.7 & ... & I* & ... & ... \\
6 & 03:29:10.49 & +31:13:30.8 & IRAS 4a & I* & ... & ... \\
7 & 03:29:11.25 & +31:18:31.7 & IRAS 6; ASR 32 & I* & ... & 0.96 \\
8 & 03:29:12.04 & +31:13:02.0 & IRAS 4b & I* & ... & 0.18 \\
9 & 03:29:13.60 & +31:13:58.2 & IRAS 4c & I* & ... & -0.07 \\
10 & 03:29:17.16 & +31:27:46.1 & ... & I* & ... & -0.53 \\
11 & 03:29:23.48 & +31:33:29.4 & IRAS 03262+3123 & I* & ... & ... \\
12 & 03:28:32.55 & +31:11:04.8 & LAL 38 & I & ... & 0.10 \\
13 & 03:28:34.49 & +31:00:51.2 & IRAS 03254+3050 & I & 0.97 & 0.64 \\
14 & 03:28:34.52 & +31:07:05.5 & ... & I & ... & 1.59 \\
15 & 03:28:37.09 & +31:13:31.1 & IRAS 1; LAL 58 & I & ... & 2.66 \\
16 & 03:28:39.69 & +31:17:32.0 & LAL 68 & I & ... & 0.34 \\
17 & 03:28:40.62 & +31:17:56.5 & ... & I & ... & 2.44 \\
18 & 03:28:43.27 & +31:17:33.1 & SVS 9; IRAS 5; ASR 126; LAL 79 & I & 0.89 & 0.52 \\
19 & 03:28:48.76 & +31:16:08.9 & ASR 67 & I & ... & 0.37 \\
20 & 03:28:51.27 & +31:17:39.5 & ASR 41; LAL 111 & I & 0.37 & 1.06 \\
21 & 03:28:55.52 & +31:14:35.8 & IRAS 2a & I & ... & 1.97 \\
22 & 03:28:57.09 & +31:21:25.1 & ... & I & ... & -0.27 \\
23 & 03:28:57.11 & +31:19:11.9 & ASR 64 & I & ... & 0.48 \\
24 & 03:28:57.99 & +31:20:52.1 & ... & I & ... & 1.32 \\
25 & 03:28:58.42 & +31:22:17.7 & LAL 166 & I & ... & 1.24 \\
26 & 03:28:59.31 & +31:20:08.2 & ASR 116; LAL 170 & I & ... & 0.65 \\
27 & 03:29:01.54 & +31:20:20.7 & SVS 12; ASR 114 & I & ... & 1.11 \\
28 & 03:29:03.32 & +31:23:14.8 & LAL 191 & I & ... & 1.94 \\
29 & 03:29:03.76 & +31:16:04.0 & SVS 13; ASR 1; LAL 196 & I & 1.01 & 1.16 \\
30 & 03:29:04.94 & +31:20:38.6 & LAL 204 & I & ... & 0.06 \\
31 & 03:29:07.77 & +31:21:57.3 & LAL 213 & I & ... & 0.56 \\
32 & 03:29:08.95 & +31:22:56.3 & LAL 221 & I & 1.79 & ... \\
33 & 03:29:09.07 & +31:21:29.1 & LAL 223 & I & ... & 1.12 \\
34 & 03:29:09.09 & +31:23:05.7 & LAL 222 & I & 0.91 & 1.07 \\
35 & 03:29:10.71 & +31:18:21.2 & ... & I & ... & 0.97 \\
36 & 03:29:12.95 & +31:18:14.6 & ASR 30; LAL 261 & I & ... & 1.03 \\
37 & 03:29:18.25 & +31:23:19.9 & ... & I & ... & 1.69 \\
38 & 03:29:22.28 & +31:13:54.5 & ... & I & ... & 0.69 \\
39 & 03:29:31.54 & +31:25:27.8 & ... & I & ... & -0.32 \\
40 & 03:28:15.19 & +31:17:23.8 & ... & II & 0.04 & -0.96 \\
41 & 03:28:34.47 & +31:17:43.2 & ... & II & ... & -0.78 \\
42 & 03:28:34.85 & +31:16:04.5 & ... & II & ... & -0.44 \\
43 & 03:28:38.76 & +31:18:06.7 & LAL 63 & II & ... & -0.12 \\
44 & 03:28:43.23 & +31:10:42.7 & LAL 80 & II & 1.51 & -1.72 \\
45 & 03:28:43.56 & +31:17:36.5 & ASR 127 & II & 0.41 & -0.24 \\
46 & 03:28:44.08 & +31:20:52.9 & LAL 83 & II & 0.22 & -1.32 \\
47 & 03:28:46.19 & +31:16:38.7 & ASR 128; LAL 89 & II & 0.27 & -1.51 \\
48 & 03:28:47.63 & +31:24:06.1 & ... & II & 0.83 & -1.61 \\
49 & 03:28:47.82 & +31:16:55.3 & SVS 17; ASR 111; LAL 97 & II & 0.27 & -1.05 \\
50 & 03:28:51.02 & +31:18:18.5 & SVS 10; ASR 122; LAL 106 & II & 0.41 & -0.62 \\
51 & 03:28:51.07 & +31:16:32.6 & ASR 44; LAL 107 & II & 0.04 & -1.09 \\
52 & 03:28:51.19 & +31:19:54.9 & ASR 125; LAL 110 & II & 0.57 & -0.83 \\
53 & 03:28:52.13 & +31:15:47.2 & ASR 45; LAL 125 & II & 0.00 & -1.17 \\
54 & 03:28:52.15 & +31:22:45.4 & LAL 120 & II & 0.30 & -1.36 \\
55 & 03:28:52.90 & +31:16:26.6 & ASR 46; LAL 128 & II & 0.00 & -1.18 \\
56 & 03:28:53.58 & +31:12:14.7 & ASR 97 & II & ... & -0.82 \\
57 & 03:28:53.93 & +31:18:09.3 & ASR 40; LAL 129 & II & 1.95 & -0.76 \\
58 & 03:28:54.07 & +31:16:54.5 & ASR 42; LAL 131 & II & 0.33 & -0.70 \\
59 & 03:28:54.61 & +31:16:51.3 & SVS 18; ASR 43; LAL 136 & II & 0.93 & -0.70 \\
60 & 03:28:54.92 & +31:15:29.2 & ASR 109; LAL 138 & II & 0.19 & -1.06 \\
61 & 03:28:55.07 & +31:16:28.8 & ASR 107; LAL 141 & II & 1.20 & -0.89 \\
62 & 03:28:55.15 & +31:16:24.8 & ASR 108 & II & 0.65 & -0.97 \\
63 & 03:28:56.09 & +31:19:08.6 & ... & II & ... & 0.47 \\
64 & 03:28:56.31 & +31:22:28.0 & LAL 147 & II & 2.19 & -0.22 \\
65 & 03:28:56.64 & +31:18:35.7 & ASR 120; LAL 150 & II & 0.80 & -1.45 \\
66 & 03:28:56.93 & +31:20:48.6 & LAL 152 & II & 0.05 & -1.17 \\
67 & 03:28:56.95 & +31:16:22.3 & SVS 15; ASR 118; LAL 154 & II & 1.30 & -0.91 \\
68 & 03:28:57.17 & +31:15:34.6 & ASR 17; LAL 156 & II & 0.70 & -0.16 \\
69 & 03:28:57.70 & +31:19:48.1 & ASR 113 & II & 0.32 & -0.63 \\
70 & 03:28:58.24 & +31:22:09.4 & LAL 164 & II & 0.67 & ... \\
71 & 03:28:58.25 & +31:22:02.1 & LAL 163 & II & 0.43 & -0.85 \\
72 & 03:28:58.43 & +31:22:56.8 & LAL 165 & II & 0.69 & -1.80 \\
73 & 03:28:59.32 & +31:15:48.5 & SVS 16; ASR 106; LAL 171 & II & 3.53 & -0.25 \\
74 & 03:28:59.56 & +31:21:46.8 & LAL 173 & II & 0.44 & -1.34 \\
75 & 03:29:00.15 & +31:21:09.3 & LAL 175 & II & 1.01 & -1.16 \\
76 & 03:29:02.16 & +31:16:11.4 & ASR 3; LAL 184 & II & 0.00 & ... \\
77 & 03:29:02.79 & +31:22:17.4 & LAL 186 & II & 1.20 & -0.87 \\
78 & 03:29:03.13 & +31:22:38.2 & LAL 189 & II & 0.37 & -0.57 \\
79 & 03:29:03.20 & +31:25:45.3 & LAL 190 & II & 0.31 & -0.96 \\
80 & 03:29:03.39 & +31:18:40.1 & ASR 63; LAL 193 & II & 0.71 & -1.79 \\
81 & 03:29:03.42 & +31:25:14.5 & LAL 192 & II & 0.94 & -1.04 \\
82 & 03:29:03.86 & +31:21:48.8 & LAL 195 & II & 0.54 & -0.87 \\
83 & 03:29:03.93 & +31:23:31.2 & ... & II & ... & -1.06 \\
84 & 03:29:04.67 & +31:16:59.2 & ASR 105; LAL 203 & II & 0.63 & -0.32 \\
85 & 03:29:04.73 & +31:11:35.0 & ASR 99; LAL 205 & II & ... & -0.34 \\
86 & 03:29:05.64 & +31:20:10.7 & ... & II & ... & -0.20 \\
87 & 03:29:05.67 & +31:21:33.8 & LAL 206 & II & ... & -0.61 \\
88 & 03:29:05.76 & +31:16:39.7 & SVS 14; ASR 7; LAL 207 & II & 2.19 & -0.28 \\
89 & 03:29:06.32 & +31:13:46.5 & ASR 53; LAL 208 & II & ... & -0.50 \\
90 & 03:29:06.92 & +31:29:57.1 & ... & II & 0.35 & -1.54 \\
91 & 03:29:07.94 & +31:22:51.6 & LAL 215 & II & 1.16 & -1.07 \\
92 & 03:29:08.95 & +31:26:24.1 & LAL 220 & II & 0.57 & -0.42 \\
93 & 03:29:09.33 & +31:21:04.2 & LAL 225 & II & 1.51 & -0.43 \\
94 & 03:29:09.47 & +31:27:21.0 & ... & II & 0.26 & -0.96 \\
95 & 03:29:09.63 & +31:22:56.5 & SVS 7; LAL 226 & II & 0.35 & -0.56 \\
96 & 03:29:10.18 & +31:27:15.9 & LAL 228 & II & 0.00 & ... \\
97 & 03:29:10.40 & +31:21:59.3 & SVS 3; LAL 230 & II & 0.62 & -0.40 \\
98 & 03:29:10.46 & +31:23:35.0 & LAL 231 & II & 1.07 & -1.02 \\
99 & 03:29:10.82 & +31:16:42.7 & ASR 23; LAL 235 & II & 0.64 & 0.09 \\
100 & 03:29:11.33 & +31:22:57.1 & LAL 239 & II & 0.66 & ... \\
101 & 03:29:11.64 & +31:20:37.7 & ASR 85; LAL 245 & II & 1.27 & -1.20 \\
102 & 03:29:11.77 & +31:26:09.7 & LAL 246 & II & ... & 0.16 \\
103 & 03:29:11.87 & +31:21:27.1 & LAL 248 & II & ... & -0.18 \\
104 & 03:29:12.90 & +31:23:29.5 & LAL 258 & II & 0.07 & -1.52 \\
105 & 03:29:13.04 & +31:17:38.4 & ASR 28; LAL 263 & II & 0.00 & -1.17 \\
106 & 03:29:13.13 & +31:22:52.9 & LAL 262 & II & 1.02 & -1.30 \\
107 & 03:29:15.34 & +31:29:34.6 & ... & II & ... & -0.39 \\
108 & 03:29:15.66 & +31:19:11.1 & ASR 59 & II & ... & -1.08 \\
109 & 03:29:16.59 & +31:23:49.6 & LAL 276 & II & 0.81 & -0.70 \\
110 & 03:29:16.81 & +31:23:25.4 & LAL 279 & II & 0.24 & -0.48 \\
111 & 03:29:17.66 & +31:22:45.2 & SVS 2; LAL 283 & II & 0.31 & -0.81 \\
112 & 03:29:17.76 & +31:19:48.2 & ASR 80; LAL 286 & II & 0.40 & -1.33 \\
113 & 03:29:18.65 & +31:20:17.9 & ASR 81 & II & ... & 0.00 \\
114 & 03:29:18.72 & +31:23:25.5 & LAL 293 & II & 0.08 & 0.08 \\
115 & 03:29:20.05 & +31:24:07.6 & LAL 296 & II & ... & 0.13 \\
116 & 03:29:20.42 & +31:18:34.3 & SVS 5; IRAS 03262+3108; ASR 112; LAL 300 & II & 1.55 & -0.60 \\
117 & 03:29:21.30 & +31:23:46.5 & LAL 302 & II & ... & -1.02 \\
118 & 03:29:21.55 & +31:21:10.5 & LAL 304 & II & 0.00 & -1.25 \\
119 & 03:29:21.87 & +31:15:36.4 & ASR 123; LAL 307 & II & 0.21 & -1.49 \\
120 & 03:29:23.15 & +31:20:30.5 & ASR 78; LAL 310 & II & 0.00 & -1.02 \\
121 & 03:29:23.23 & +31:26:53.1 & LAL 308 & II & 0.16 & -1.02 \\
122 & 03:29:24.08 & +31:19:57.8 & ASR 79; LAL 313 & II & 0.58 & -0.39 \\
123 & 03:29:25.92 & +31:26:40.1 & SVS 4; LAL 318 & II & ... & 0.06 \\
124 & 03:29:28.01 & +31:25:11.0 & ... & II & ... & -1.00 \\
125 & 03:29:29.79 & +31:21:02.8 & LAL 333 & II & 0.39 & -0.81 \\
126 & 03:29:30.39 & +31:19:03.4 & LAL 336 & II & 0.00 & -1.05 \\
127 & 03:29:32.56 & +31:24:37.0 & LAL 342 & II & 0.38 & -1.43 \\
128 & 03:29:32.86 & +31:27:12.7 & LAL 344 & II & 0.08 & -1.67 \\
129 & 03:29:35.70 & +31:21:08.6 & ... & II & ... & -0.27 \\
130 & 03:29:37.63 & +31:02:49.3 & ... & II & 0.00 & -1.14 \\
131 & 03:29:37.72 & +31:22:02.6 & ... & II & 0.00 & -1.29 \\
132 & 03:29:44.15 & +31:19:47.5 & ... & II & ... & -0.64 \\
133 & 03:29:54.03 & +31:20:53.1 & ... & II & 0.00 & -0.99 \\
134 & 03:29:02.69 & +31:19:05.8 & ASR 62; LAL 187 & II/III & ... & -1.71 \\
135 & 03:29:09.41 & +31:14:14.1 & ASR 54 & II/III & ... & -0.32 \\
136 & 03:29:26.79 & +31:26:47.7 & SVS 6; LAL 321 & II/III & 0.22 & -2.14 \\
137 & 03:29:29.26 & +31:18:34.8 & LAL 331 & II/III & 0.65 & -2.12 \\
\enddata
\tablenotetext{a}{J2000}
\tablenotetext{b}{Asterisks mark "deeply embedded" sources with questionable IRAC colors or incomplete IRAC photometry and relatively bright MIPS 24~$\mu$m photometry.}
\tablenotetext{c}{Only provided for sources with valid $JHK_S$ photometry.}
\tablenotetext{d}{Extinction is not accounted for in these values.  High extinction can bias $\alpha_{IRAC}$ to higher values.}
\end{deluxetable}

\begin{deluxetable}{ccccccc}
\tabletypesize{\scriptsize}
\tablecaption{Summary of Disk Fraction Measurements by Region\label{dftable}}
\tablewidth{0pt}
\tablehead{
\colhead{Region} & \colhead{R.A.\tablenotemark{a}} & \colhead{Decl.\tablenotemark{a}} & \colhead{Radius (arcsec)} & \colhead{No. with IR-excess} & \colhead{Total No. of Members} & \colhead{Disk Fraction}
}
\startdata
Main   & 03:29:04.67 & +31:19:57.9 & 330 &   72    & 87   & $0.83 \pm 0.11$ \\
S1     & 03:28:55.12 & +31:16:23.8 &  90 &   11    & 15   & $0.73 \pm 0.07$ \\
S2     & 03:29:03.47 & +31:16:54.4 &  90 &    6    &  7   & $0.86 \pm 0.10$ \\
N1     & 03:29:10.64 & +31:23:01.4 &  90 &   13    & 14   & $0.93 \pm 0.07$ \\
N2     & 03:29:02.28 & +31:21:29.7 &  90 &   14    & 16   & $0.88 \pm 0.07$ \\
E      & 03:29:27.34 & +31:19:57.7 & 120 &   7     & 7    & $1.00 \pm 0.13$ \\
\enddata
\tablenotetext{a}{J2000}
\end{deluxetable}

\clearpage

\begin{figure}
\epsscale{1}
\plotone{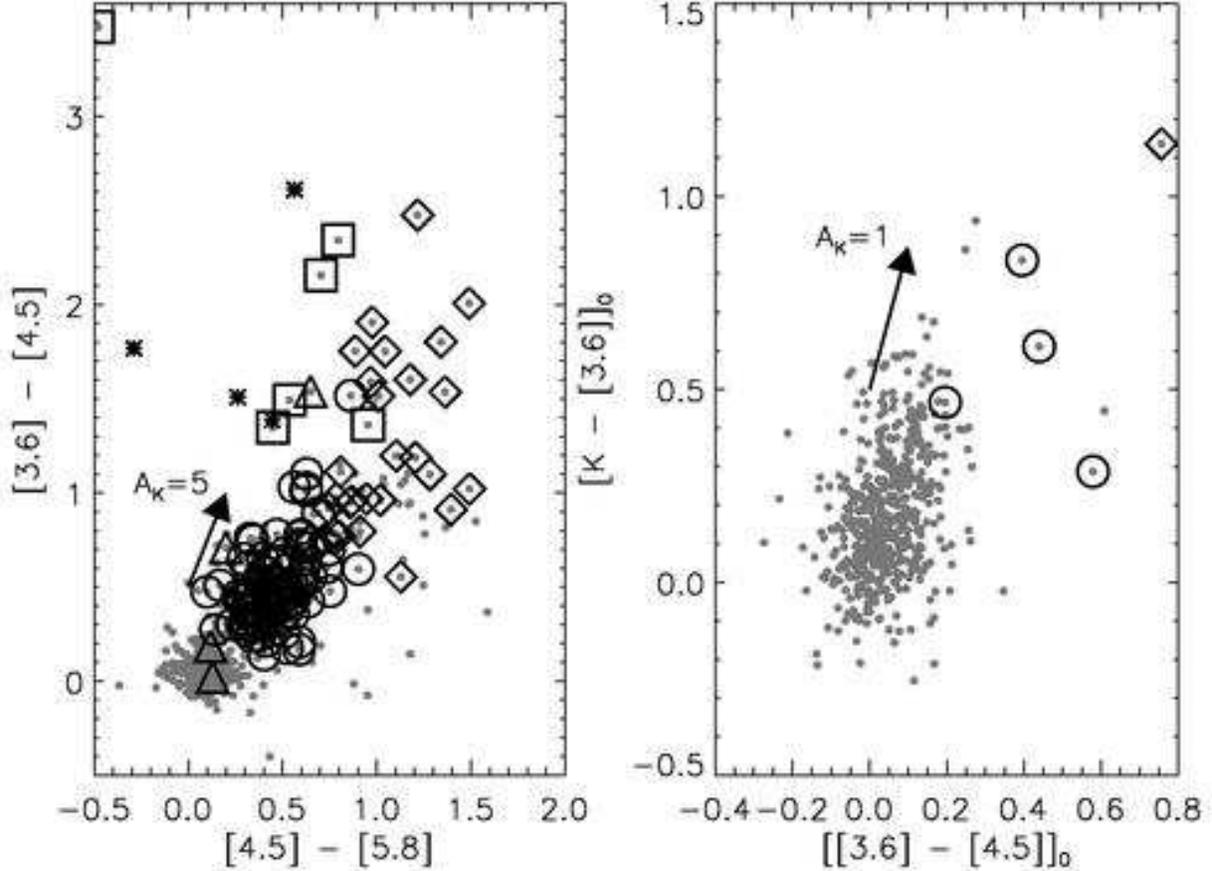}
\caption{Color-color diagrams used for YSO identification and classification.  The left plot is the $[3.6]-[4.5]$ vs. $[4.5]-[5.8]$ color-color diagram, used for identifying likely protostars and filtering out barely-resolved shock emission.  Diamonds mark the IRAC-identified protostellar sources, and circles mark the more-evolved Class~II sources.  Triangles mark transition/debris disk candidates, sources with colors consistent with reddened photospheres that have IR-excess at 24~$\mu$m.  Asterisks mark likely shock emission knots.  Squares mark ``deeply embedded YSOs'', likely protostars, that are composed of shock knot and broad-line AGN candidates and sources with incomplete IRAC four-band photometry that have particularly red IRAC/MIPS colors. The right plot is the $[K-[3.6]]_0$ vs. $[[3.6]-[4.5]]_0$ color-color diagram, used for secondary YSO identification and classification for sources that lack either 5.8 or 8.0~$\mu$m photometry.\label{proto1}}
\end{figure}

\begin{figure}
\epsscale{.5}
\plotone{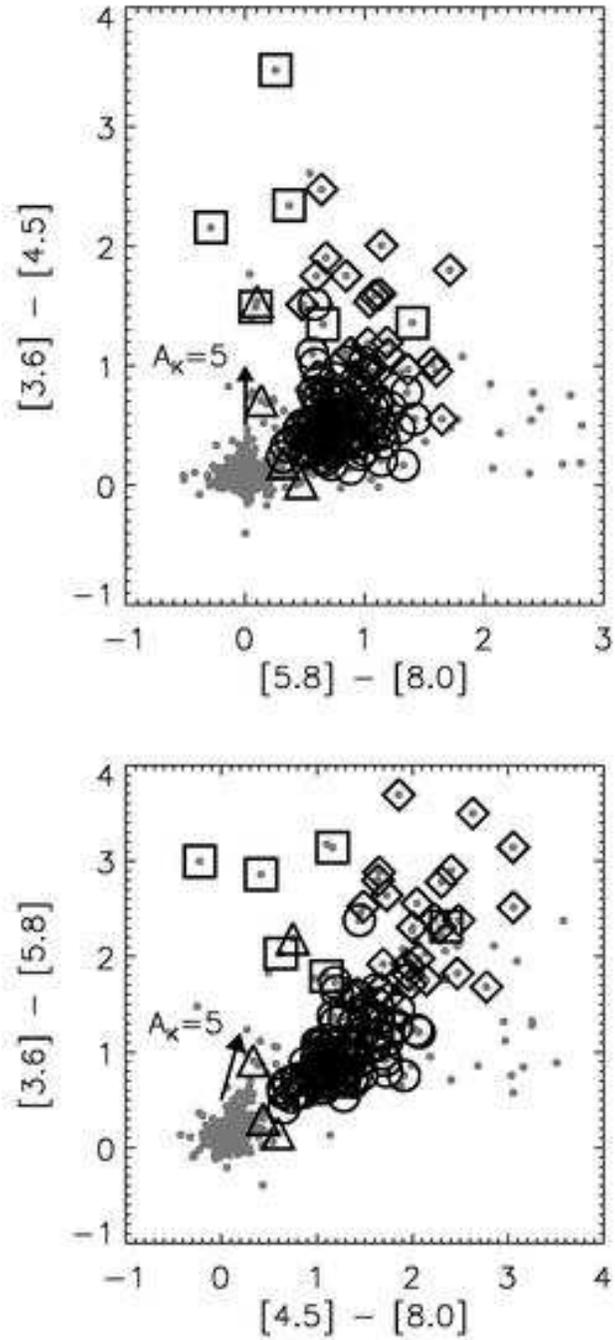}
\caption{Two IRAC four-band color-color diagrams used for classifying YSOs.  The top plot is the $[3.6]-[4.5]$ vs. $[5.8]-[8.0]$ color-color diagram, used in several previous papers \citep[e.g.][]{mege04,alle04}.  The bottom plot is the $[3.6]-[5.8]$ vs. $[4.5]-[8.0]$ color-color diagram, used in this work for identifying likely Class~II sources.  Plot overlays are the same as Figure~\ref{proto1}.\label{ctts1}}
\end{figure}


\begin{figure}
\epsscale{1}
\plotone{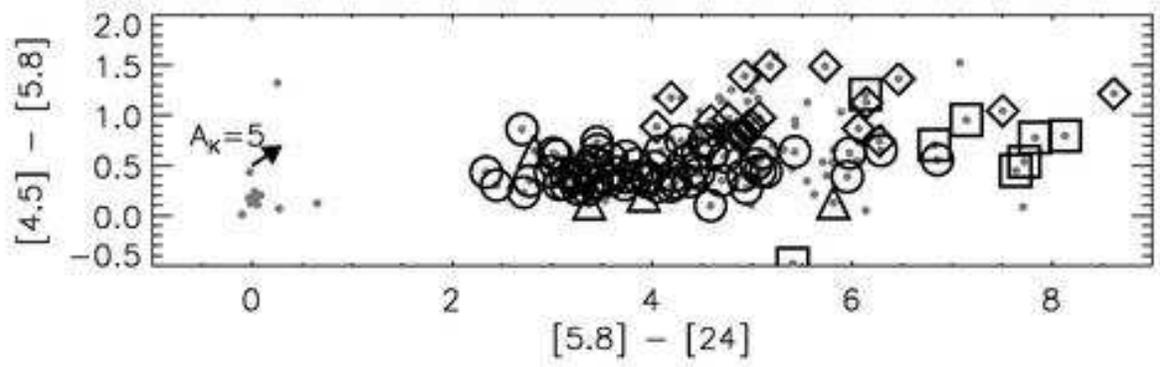}
\caption{$[4.5]-[5.8]$ vs. $[5.8]-[24]$ color-color diagram, used for identifying likely transition/debris disk sources and to verify protostellar classifications.  Plot overlays are the same as Figure~\ref{proto1}.\label{mipscheck}}
\end{figure}


\clearpage


\begin{figure}
\epsscale{1}
\plotone{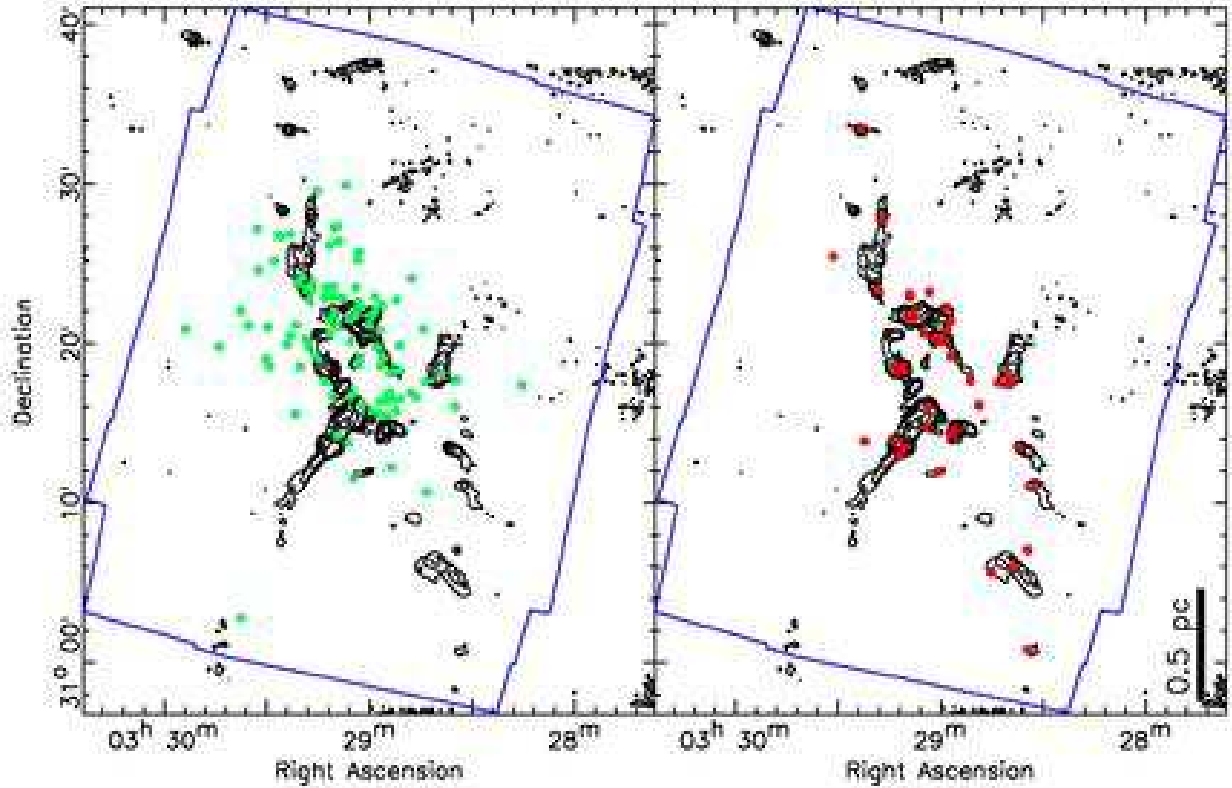}
\caption{SCUBA 850~$\mu$m contours with Class~II sources overlaid on the left in green and Class~I sources on the right in red.  The first contour level marks 0.05 Jy/beam, and each successive contour is double the previous level.  Note that the Class~I population traces the dense material much more closely than the Class~II sources.\label{scubadistro}}
\end{figure}

\begin{figure}
\epsscale{1}
\plotone{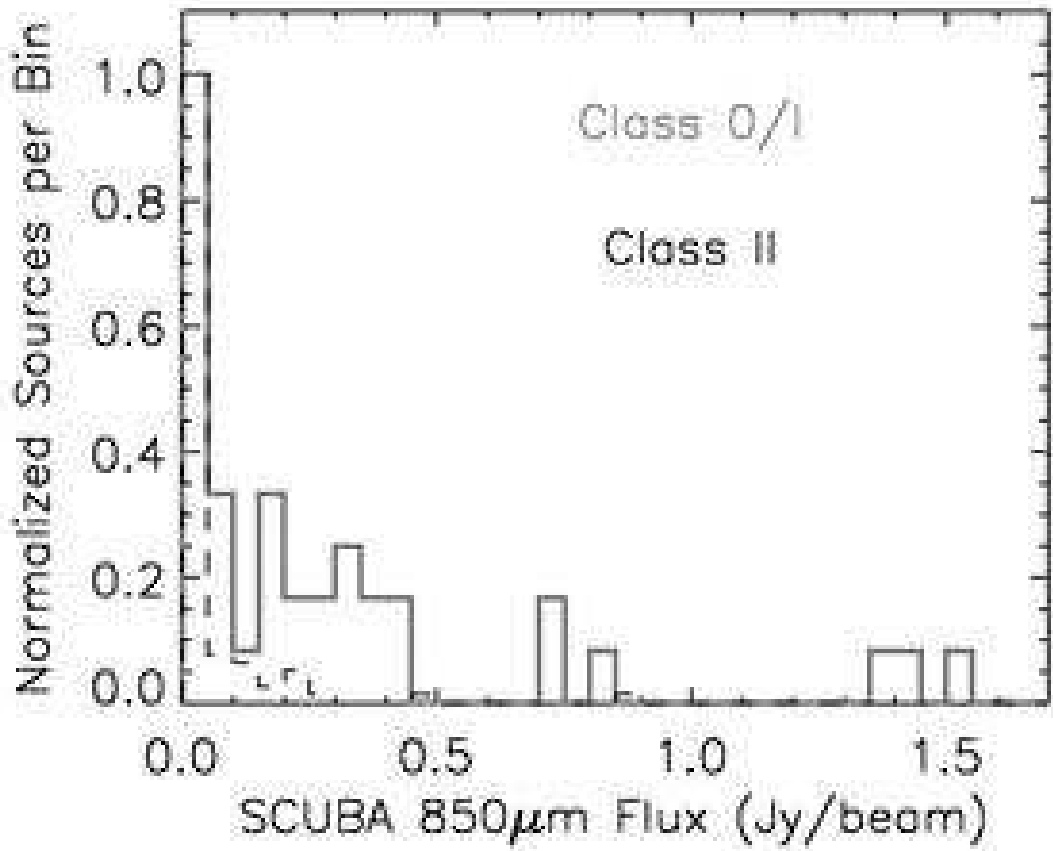}
\caption{Normalized histograms comparing the SCUBA 850~$\mu$m flux at the positions of Class~I and Class~II populations in gray and black, respectively.  This figure demonstrates that protostars are much more strongly confined to regions of high gas density compared to their more evolved counterparts.\label{1v2cdf}}
\end{figure}

\begin{figure}
\epsscale{.7}
\plotone{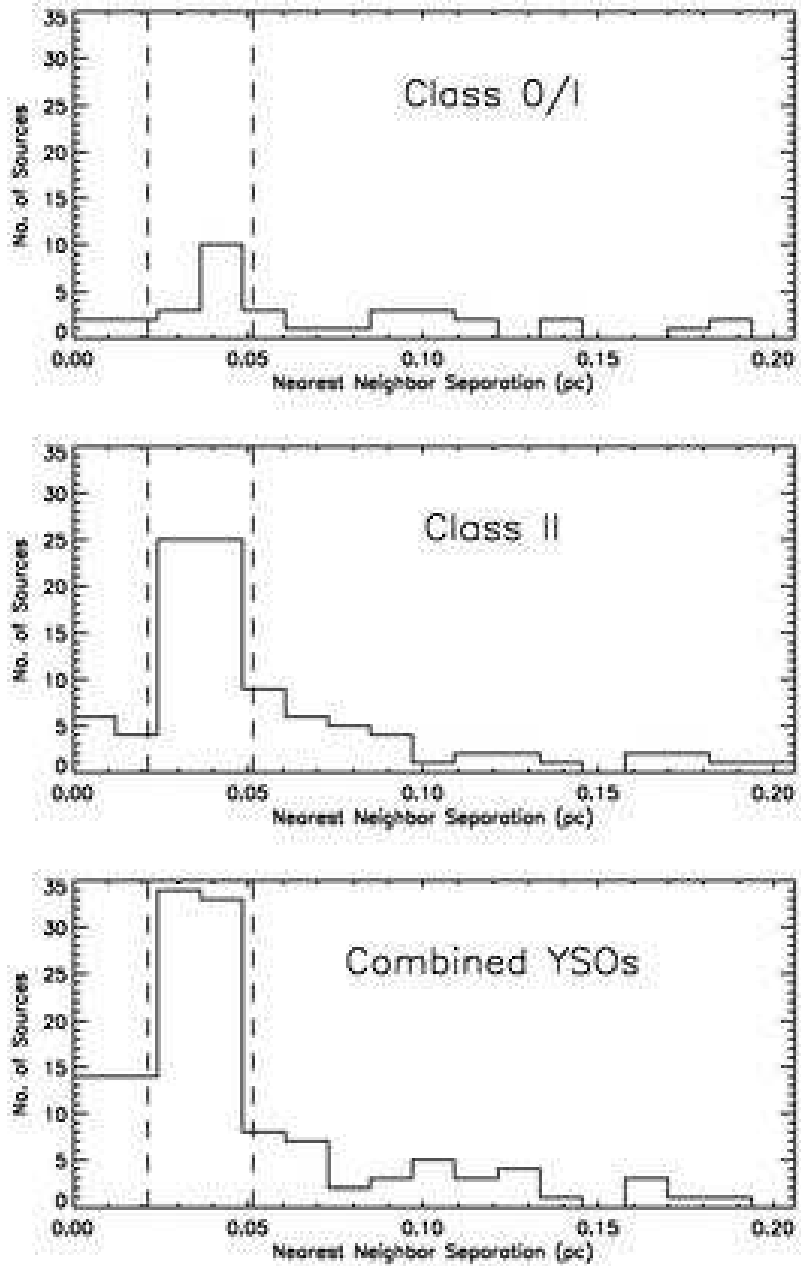}
\caption{Histograms of nearest neighbor distances in NGC~1333.  Top and middle histograms are the distributions among the two evolutionary classes, protostars and Class~II respectively.  The bottom histogram is the distribution for all YSOs together.  The dashed vertical lines mark the range where all three distributions have clear peaks.\label{nnd_hist}}
\end{figure}

\begin{figure}
\epsscale{1}
\plotone{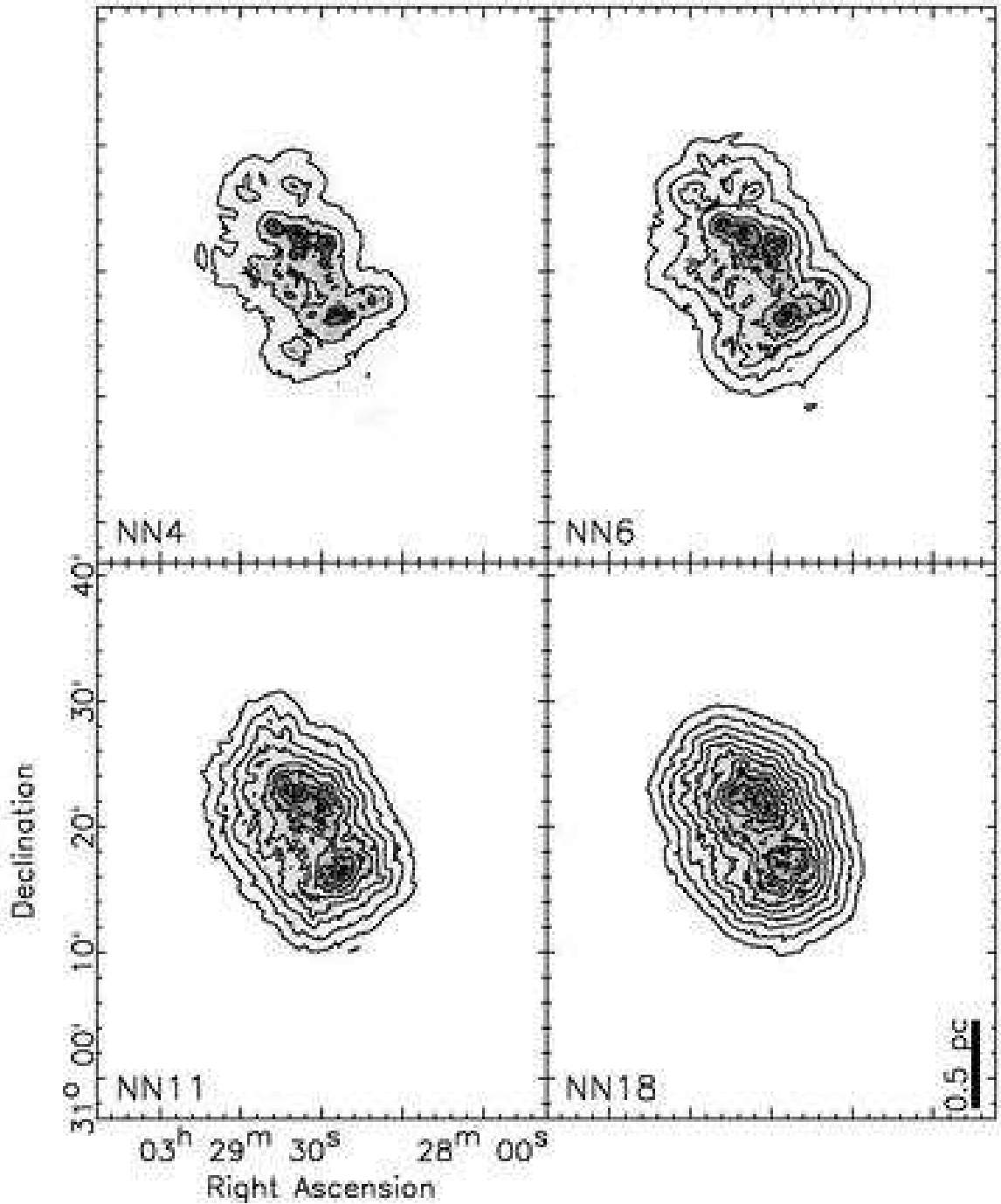}
\caption{A comparison of NN$n$ surface density mapping results for the NGC~1333 YSOs, using $n$ = 4, 6, 11, and 18 neighbors.  As the number of neighbors used is increased, smaller scale structure is overwhelmed by the large scale distribution that is dominated by the ``double cluster''.  Contours levels mark stellar surface densities of 1~$\sigma$ below successive contours \citep{ch85}, and the linear inverse grayscale sets 1000~pc$^{-2}$ as black and 0~pc$^{-2}$ as white.\label{nngs}}
\end{figure}

  
\begin{figure}
\epsscale{1}
\plotone{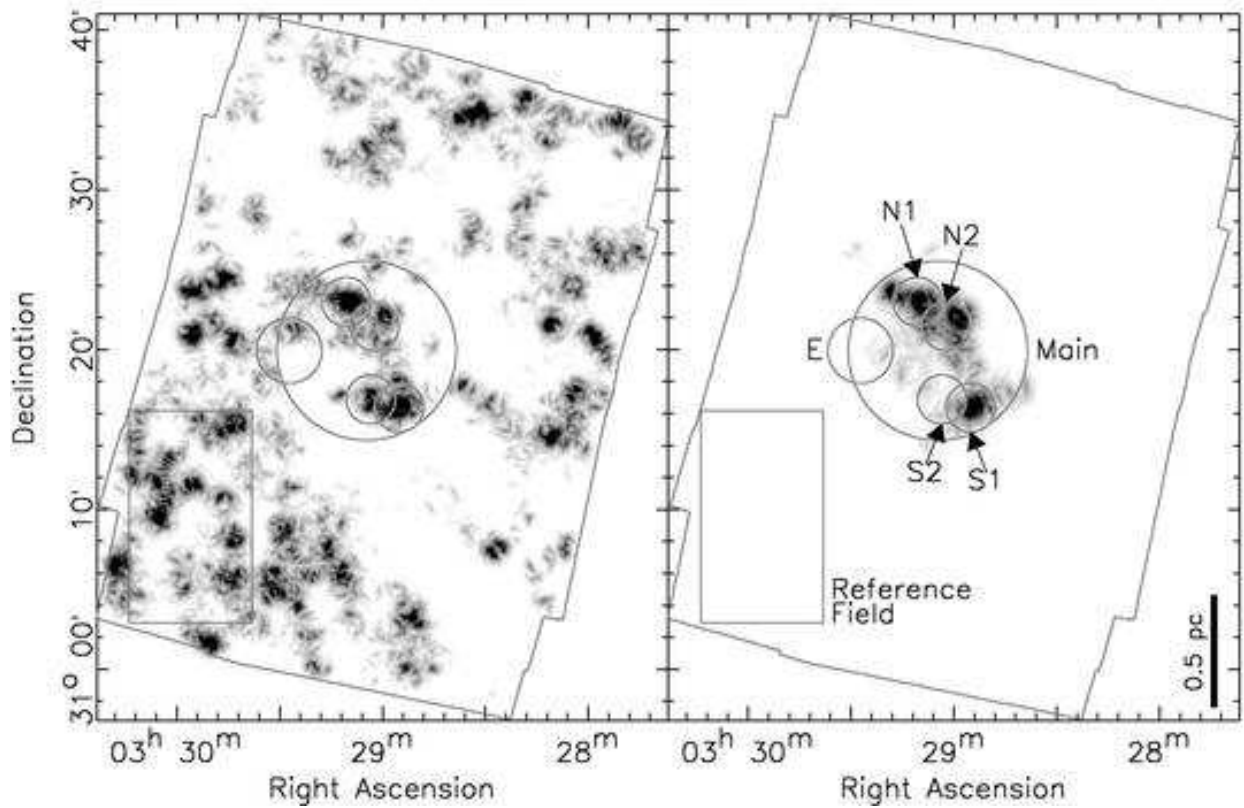}
\caption{At left: NN6 map of the surface density of all sources detected at $K_S$, $[3.6]$ and $[4.5]$~$\mu$m with $K_S < 15$.  At right: NN6 map of the surface density of {\it Spitzer}-identified YSOs only, i.e. only confirmed cluster members.  The circle labeled ``Main'' marks the cluster region used for measuring the fraction of members with circumstellar material.  Other smaller regions are also examined and their results are listed in Table~\ref{dftable}.  The box marks the control field used for characterizing field star contamination in the statistically inferred Class~III diskless pre-main sequence star population.\label{dfregion}}
\end{figure}
  
\begin{figure}
\epsscale{.6}
\plotone{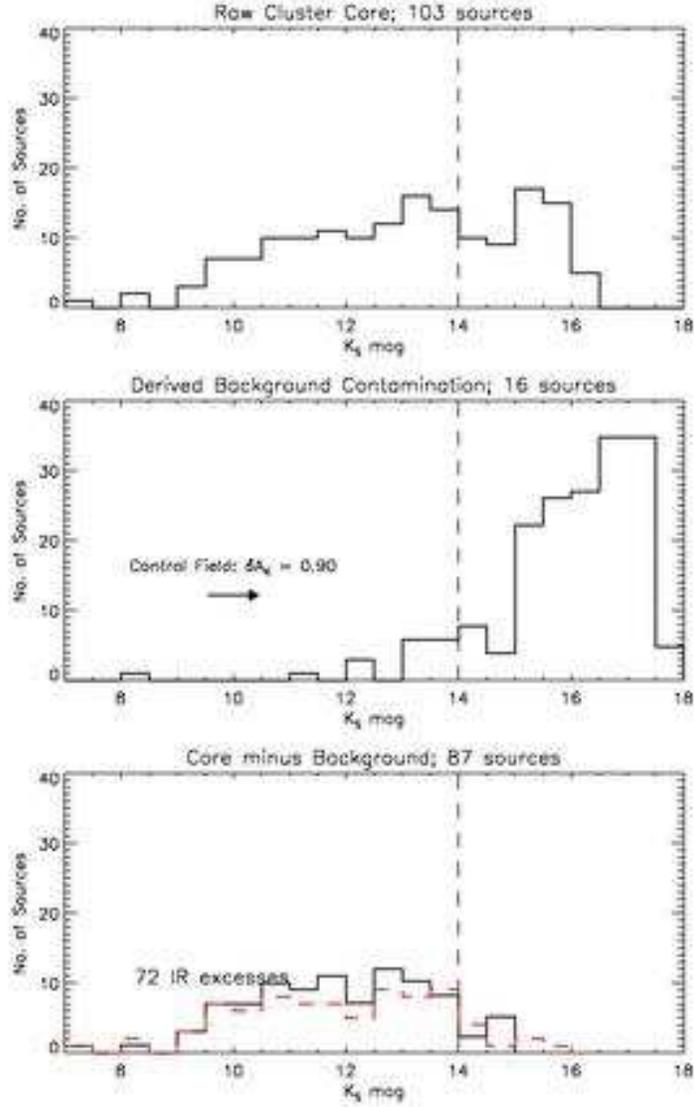}
\caption{$K_S$ magnitude histogram (KMH) demonstrating the measurement of the total membership of the cluster region in Figure~\ref{dfregion}.  The top plot shows the measured KMH for those sources within the circle that are detected at $H$, $K_S$, $[3.6]$ and $[4.5]$~$\mu$m with $K_S < 14$.  The middle plot shows the KMH of the control field, scaled by the ratio of the areas of the cluster region and the control field and shifted to account for the mean extinction difference between the two fields.  The bottom plot shows the final field star corrected cluster KMH for the cluster region, resulting in a total membership estimate of 82.  Overplotted in red is the KMH for all {\it Spitzer}-identified YSOs detected at $K_S$ band in this region.  There are 72 with $K_S < 14$, thus the fraction of members with disks is 83\%.\label{dfhist}}
\end{figure}

\begin{figure}
\epsscale{.6}
\plotone{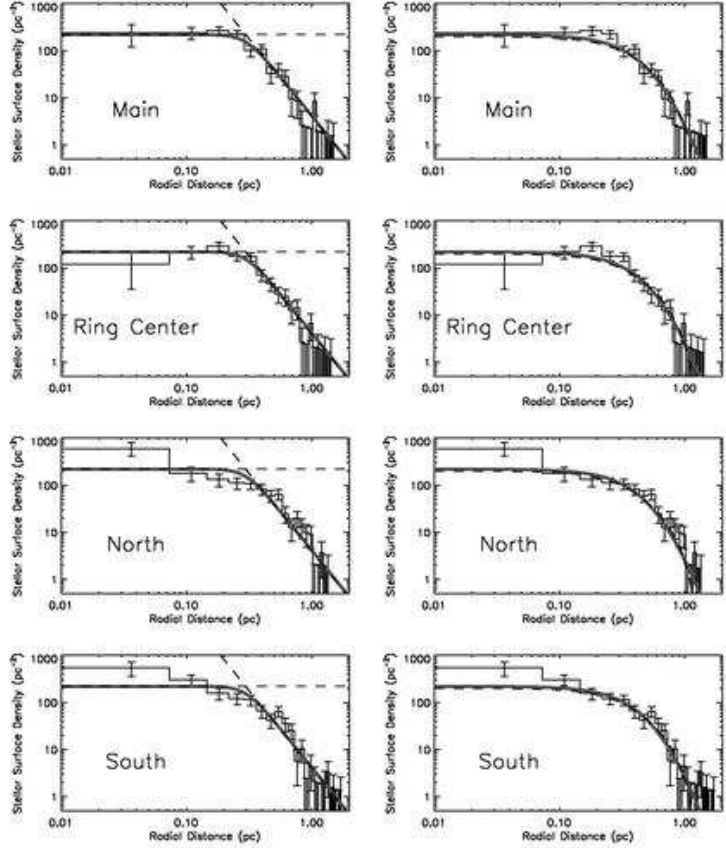}
\caption{Stellar surface density as a function of radius for the {\it Spitzer}-identified YSOs centered on four distinct choices of cluster center points.  In the left column of plots, the straight, black, dashed lines are power laws of $\alpha=0$ (shallow slope) and $\alpha=-3.3$ (steep slope) that were chosen to fit the functional behavior of the radial profiles at small and large radii, respectively.  Also in the left column, a function of the form $\sigma = \sigma_0 (1+(\frac{r}{r_0})^\beta)^{\frac{-\gamma}{\beta}}$ is plotted in solid gray, with parameters $\sigma_0 = 220$~pc$^{-2}$, $\beta=6$, $\gamma=3.3$, and $r_0 = 0.3$~pc, the location of the knee in the two power law model.  In the right column, the overlays are the best fit King profile in dashed black and the best fit EFF profile in solid gray for the ``Main'' radial profile.\label{radprof1}}
\end{figure}



\begin{figure}
\epsscale{1}
\plotone{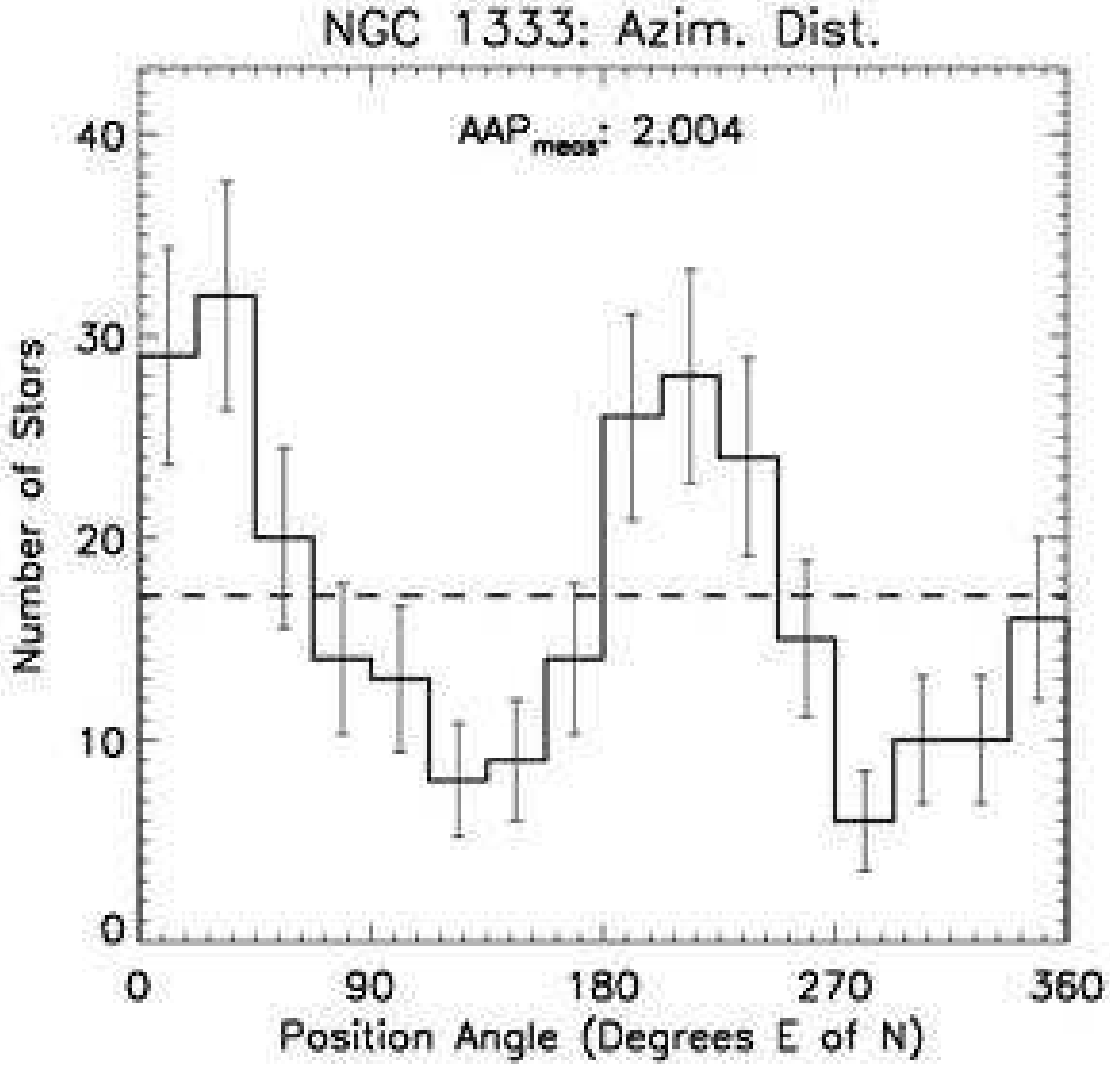}
\caption{YSO number counts per $45^{\circ}$ azimuthal bin (Nyquist sampled), with error bars derived from Poisson counting statistics for the contents of each bin.  The dashed line marks the mean source count per bin.  The AAP is the standard deviation of the values for each bin about this mean, divided by the square root of the mean.  The resulting AAP of 2 for NGC~1333 is quite high in reference to other young clusters analyzed this way \citep{gute05}.  The substructure that is dominating this measurement is likely the two-peaked nature of the ``double cluster''.\label{azhist1}}
\end{figure}

\clearpage

\begin{figure}
\epsscale{.6}
\plotone{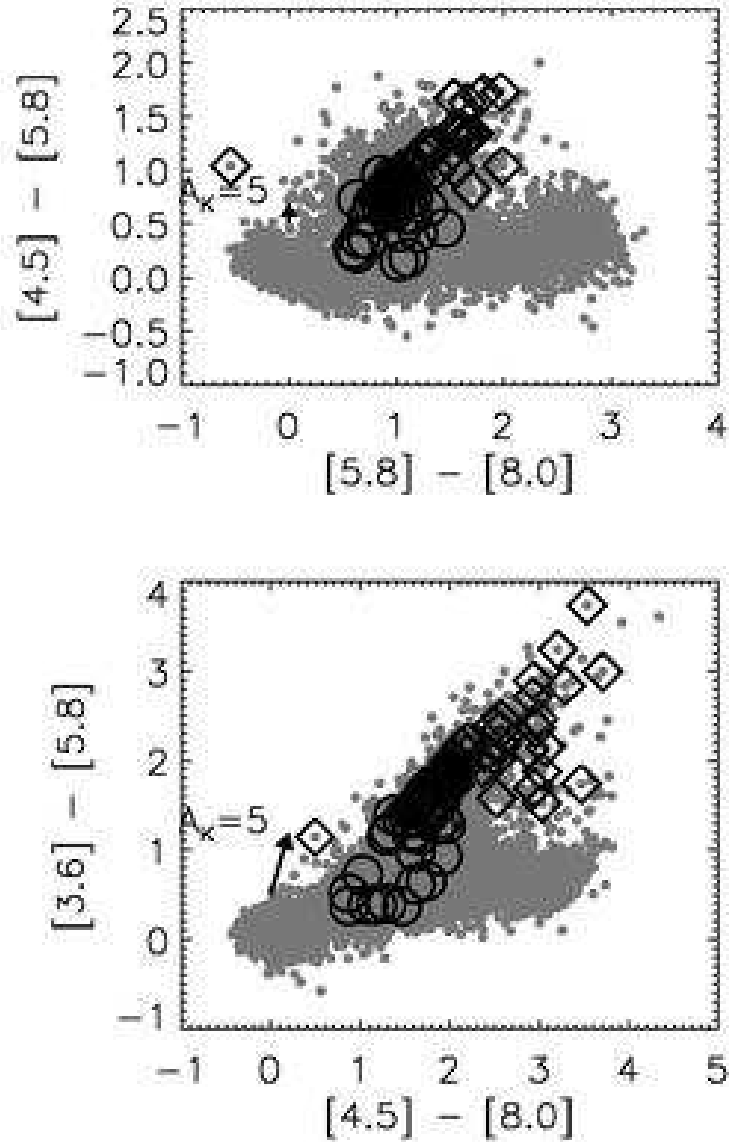}
\caption{Color-color diagrams of the Bootes field data used to identify PAH emission sources.  The top plot is the $[4.5]-[5.8]$ vs. $[5.8]-[8.0]$ color-color diagram, and the botton one is the $[3.6]-[5.8]$ vs. $[4.5]-[8.0]$ diagram.  Diamonds mark residual contaminants that have mid-IR colors and magnitudes similar to Class~I YSOs.  Circles mark residual contaminants with Class~II colors and magnitudes.\label{bpah1}}
\end{figure}


\begin{figure}
\epsscale{1}
\plotone{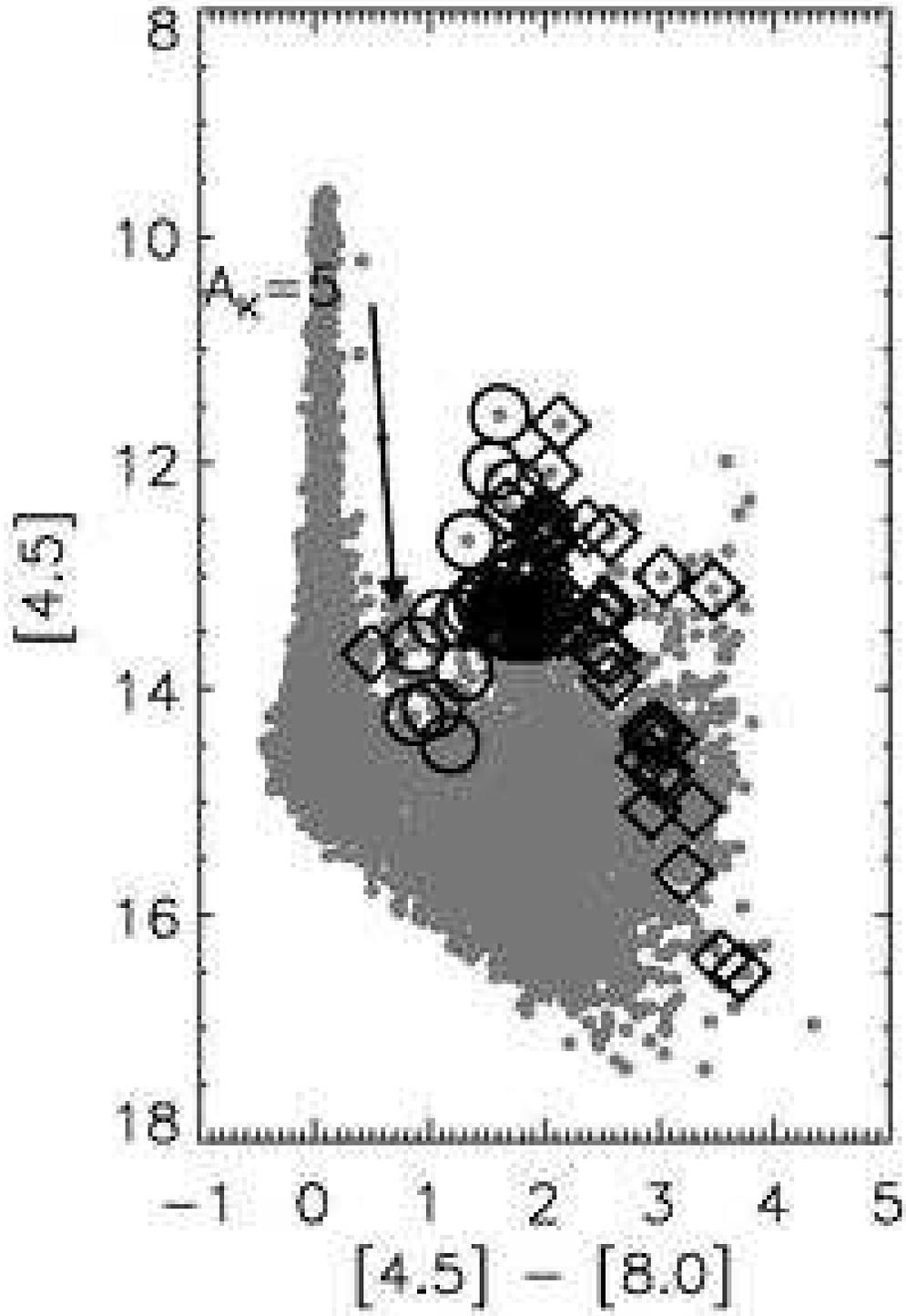}
\caption{$[4.5]$ vs. $[4.5]-[8.0]$ color-magnitude diagram, used for identifying likely broad-line AGN.  Diamonds mark residual contaminants that have mid-IR colors and magnitudes similar to Class~I YSOs.  Circles mark residual contaminants with Class~II colors and magnitudes.\label{bagn1}}
\end{figure}

\begin{figure}
\epsscale{1}
\plotone{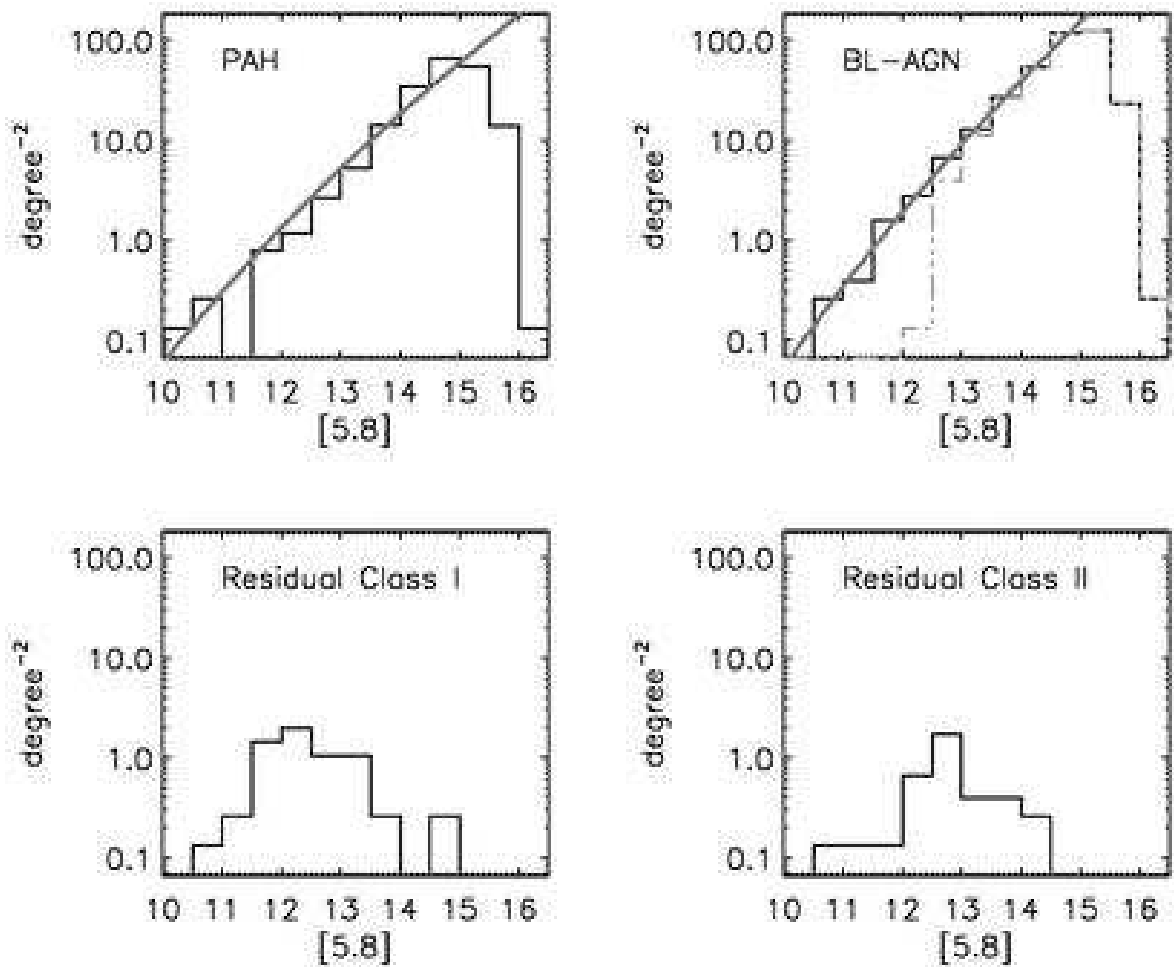}
\caption{Source counts per square degree vs. $[5.8]$ for the Bootes field.  The upper left histogram is the distribution of PAH emission sources.  The upper right solid black histogram is all other sources with IR-excess emission in the Bootes field, assumed to be dominated by broad-line AGN.  The gray dot-dashed histrogram is the distribution of sources that are successfully filtered with our broad-line AGN cut shown in Figure~\ref{bagn1}.  Gray solid straight lines in both plots are the power law fits presented in the text.  The residual contaminating sources are classed as they would be if they were YSOs to gauge contamination by class.  The lower left histrogram is for residual ``protostar'' contaminants, while the lower right is for residual ``Class~II'' contaminants.\label{bhist1}} 
\end{figure}




\end{document}